\documentclass[journal]{IEEEtran}

\usepackage{array,url}
\usepackage{amssymb,amsfonts,amsmath,amsthm,amscd,mathrsfs,mathtools,color,bbm}
\usepackage{graphicx,float,epsfig,enumerate,fixmath,latexsym,cite,booktabs,makecell}
\usepackage{tikz}
\usepackage{pgfplots}
\pgfplotsset{compat=1.18}
\usepackage[hidelinks]{hyperref}
\usepackage{xcolor}
\hypersetup{colorlinks=true,linkcolor=blue!70!black,citecolor=green!40!black,urlcolor=red!70!black}
\usetikzlibrary{shapes.geometric,arrows.meta,positioning,calc,fit}

\newcommand{\bbC}{\mathbb{C}}\newcommand{\rmc}{\mathrm{c}}
\newcommand{\rmd}{\mathrm{d}}
\newcommand{\bbE}{\mathbb{E}}\newcommand{\rme}{\mathrm{e}}

\newcommand{\bbP}{\mathbb{P}}

\newcommand{\bbR}{\mathbb{R}}
\newcommand{\rms}{\mathrm{s}}

\newcommand{\bfA}{\mathbf{A}}\newcommand{\bfa}{\mathbf{a}}
\newcommand{\bfB}{\mathbf{B}}
\newcommand{\bfc}{\mathbf{c}}

\newcommand{\bfG}{\mathbf{G}}\newcommand{\bfg}{\mathbf{g}}
\newcommand{\bfH}{\mathbf{H}}\newcommand{\bfh}{\mathbf{h}}\newcommand{\sfH}{\mathsf{H}}
\newcommand{\bfI}{\mathbf{I}}
\newcommand{\bfJ}{\mathbf{J}}

\newcommand{\bfp}{\mathbf{p}}
\newcommand{\bfQ}{\mathbf{Q}}
\newcommand{\sfR}{\mathsf{R}}
\newcommand{\bfS}{\mathbf{S}}\newcommand{\bfs}{\mathbf{s}}
\newcommand{\sfT}{\mathsf{T}}
\newcommand{\bfu}{\mathbf{u}}\newcommand{\sfU}{\mathsf{U}}
\newcommand{\bfV}{\mathbf{V}}\newcommand{\bfv}{\mathbf{v}}\newcommand{\sfV}{\mathsf{V}}
\newcommand{\sfW}{\mathsf{W}}
\newcommand{\bfX}{\mathbf{X}}\newcommand{\bfx}{\mathbf{x}}\newcommand{\sfX}{\mathsf{X}}
\newcommand{\bfY}{\mathbf{Y}}\newcommand{\sfY}{\mathsf{Y}}
\newcommand{\bfZ}{\mathbf{Z}}\newcommand{\sfZ}{\mathsf{Z}}

\newcommand{\cA}{\mathcal{A}}

\newcommand{\cC}{\mathcal{C}}

\newcommand{\cF}{\mathcal{F}}

\newcommand{\cH}{\mathcal{H}}

\newcommand{\cN}{\mathcal{N}}

\newcommand{\cP}{\mathcal{P}}
\newcommand{\cQ}{\mathcal{Q}}
\newcommand{\cR}{\mathcal{R}}

\newcommand{\sfu}{\mathsf{u}}
\newcommand{\sfv}{\mathsf{v}}\newcommand{\cV}{\mathcal{V}}
\newcommand{\sfw}{\mathsf{w}}
\newcommand{\cX}{\mathcal{X}}
\newcommand{\sfy}{\mathsf{y}}\newcommand{\cY}{\mathcal{Y}}

\newtheoremstyle{mystyle}% name
{}% space above
{}% space below
{\itshape}% body font
{}% indent amount
{\bfseries}% theorem head font
{}% punctuation after theorem head
{.5em}% space after theorem head
{}% theorem head spec

\newtheoremstyle{remark}% name
{}% space above
{}% space below
{}% body font
{}% indent amount
{\itshape}% theorem head font
{}% punctuation after theorem head
{.5em}% space after theorem head
{}% theorem head spec

\makeatletter
\def\thmhead@plain#1#2#3{%
  \thmname{#1}\thmnumber{\@ifnotempty{#1}{ }\@upn{#2.}}%
  \thmnote{ \textsf{\the\thm@notefont\textit{#3}.}}}
\let\thmhead\thmhead@plain
\makeatother

\theoremstyle{mystyle}
\newtheorem{theorem}{Theorem}%[section]
\theoremstyle{mystyle}
\newtheorem{lemma}{Lemma}%[section]
\theoremstyle{mystyle}
\newtheorem{prop}{Proposition}%[section]
\theoremstyle{mystyle}
\newtheorem{corollary}{Corollary}%[thm]
\theoremstyle{mystyle}
\newtheorem{definition}{Definition}%[section]
\theoremstyle{remark}
\newtheorem{rem}{Remark}%[section]
\theoremstyle{mystyle}
\theoremstyle{mystyle}
\theoremstyle{mystyle}
\theoremstyle{mystyle}
\theoremstyle{mystyle}
\theoremstyle{mystyle}
\usepackage{tocloft} % THIS PACKAGE DOES NOT WORK WITH \usepackage{subfigure} 
\newcommand{\independent}{\mathrel{\perp\!\!\!\perp}}

\DeclareMathOperator*{\argmin}{arg\,min}

\makeatletter
\newcommand{\vast}{\bBigg@{4}}
\newcommand{\Vast}{\bBigg@{5}}
\newcommand{\Gigantic}{\bBigg@{8}}

\makeatother
\allowdisplaybreaks

\title{A Rate-Distortion Bound for Integrated Sensing and Communication}

\author{Mohammadreza Bakhshizadeh Mohajer, Alex Dytso, Daniela Tuninetti, and Luca Barletta%
\thanks{Mohammadreza Bakhshizadeh Mohajer and Luca Barletta are with Politecnico di Milano, Milan, 20133 Italy (e-mail: \{mohammadreza.bakhshizadeh, luca.barletta\}@polimi.it). Alex Dytso is with Qualcomm Flarion Technologies, Bridgewater, NJ 08807 USA (e-mail: odytso2@gmail.com). Daniela Tuninetti is with the University of Illinois Chicago, Chicago, IL 60607 USA (e-mail: danielat@uic.edu). Her work was partially funded by NSF Award 2312229.}}

\begin{document}
\maketitle

\begin{abstract}
We develop a general rate--distortion bound (RDB) for the fundamental sensing--communication tradeoff in integrated sensing and communication (ISAC) systems. The RDB lower-bounds the Bayesian sensing risk for arbitrary parameter alphabets, sensing-channel laws, and distortion measures without requiring regularity conditions on the parameter distribution. A key feature is that the distortion--rate function is applied to the sensing information induced by each transmitted signal before averaging. This preserves signal-dependent sensing information and yields a bound that is no weaker, and potentially strictly tighter, than one based only on average sensing information. The RDB is exact in the high-sensing-noise limit and, under scalar squared-error loss, its entropy-power specialization is no looser than the Bayesian Cram\'er--Rao bound (BCRB). Applications to binary occupancy detection and multiple-input multiple-output (MIMO) Nakagami target-response estimation demonstrate its scope, including discrete sensing and severe fading regimes in which the conventional BCRB yields only the trivial zero lower bound. Covariance-based outer bounds and Gaussian and semi-unitary achievable schemes further reveal how signal randomness and spatial energy allocation shape the tradeoff.
\end{abstract}


\section{Introduction}
\label{sec:intro}
Integrated Sensing and Communications (ISAC), the joint design of wireless systems for both sensing and data transmission, has become a key technology for 6G networks. By sharing hardware and spectrum, ISAC improves energy, spectral, and hardware efficiency~\cite{ISAC_Caire,Liu2022JSAC,isac_6g_multifunc}. These gains support simultaneous sensing and communication in automotive systems (e.g., 6G V2X~\cite{V2X_mizmizi}), industrial environments (e.g., smart manufacturing), and urban environments such as smart cities, transportation systems, and homes~\cite{smartcity}. Recent surveys highlight a growing range of system designs and applications, underscoring the need for principled performance limits~\cite{6g_isac}.

From the sensing perspective, the goal is to infer unknown parameters from the return of a transmitted waveform; from the communication perspective, the goal is to reliably convey information at the highest possible rate. Since the same transmit waveform affects both tasks, their different objectives lead to a fundamental sensing--communication tradeoff. Rigorous analysis therefore requires combining information- and estimation-theoretic tools to characterize the attainable performance region~\cite{ISAC_Caire}.

\subsection{Prior Work}
Recent works have studied ISAC from signal-processing, information-theoretic, and waveform-design perspectives~\cite{Liu2022JSAC,isac_6g_multifunc,6g_isac}. Existing approaches, however, typically specialize to particular state models, sensing criteria, or regularity assumptions; a single converse framework that directly accommodates continuous, discrete, and hybrid sensing parameters together with general distortion measures is less common.

A closely related information-theoretic formulation is the capacity--distortion--cost framework of Ahmadipour \emph{et al.}~\cite{ahmadipour2024information}. For a single-receiver state-dependent memoryless channel with an i.i.d. state sequence, their~\cite[Theorem~1]{ahmadipour2024information} characterizes the capacity--distortion--cost tradeoff by optimizing communication mutual information over input laws satisfying both an estimation-distortion constraint~\cite[eq(9a)]{ahmadipour2024information} and an input-cost constraint~\cite[eq(9b)]{ahmadipour2024information}. They further provide a modified Blahut--Arimoto procedure for numerical evaluation in discrete models. Our setting differs operationally in that one sensing parameter is inferred separately within each coherence block of $T$ channel uses, whereas the communication code spans many independent coherence blocks, consistently with the blockwise formulation in~\cite{ISAC_Caire}. Thus, sensing cannot exploit coding across independent parameter realizations in the same manner as the communication code. Our objective is also different: rather than exactly solving a capacity--distortion problem for a prescribed memoryless state model, we seek a broadly applicable outer bound whose sensing term can be expressed and subsequently relaxed through information-theoretic quantities. This does not remove the input-distribution optimization, but it exposes structure that can be exploited through covariance and other problem-specific relaxations.

For Gaussian ISAC channels, the Bayesian Cram\'er--Rao bound (BCRB) has been used to lower-bound the minimum mean-squared error (MMSE) and characterize sensing--rate tradeoffs~\cite{ISAC_Caire,HuaHanXu2024,miller1978modified}. Other estimation-theoretic lower bounds include the Weiss--Weinstein family~\cite{WWbound}, the Ziv--Zakai family~\cite{ziv1969some,Ziv-ZakBOund}, bounds from functional inequalities, including log-Sobolev inequalities~\cite{dytso2023meta,aras2019family}, and bounds based on variational divergence representations~\cite{dytso2019mmse}.

In this work, we use the rate--distortion bound (RDB), an information-theoretic lower bound on Bayesian risk that has received limited attention in estimation theory. The underlying connection to rate--distortion (RD) theory can be traced at least to the early work of Goblick~\cite{goblick1965theoretical}. The RDB applies to any well-defined loss function, thereby covering sensing criteria beyond MSE, including average probability of error. Unlike the BCRB, it does not require regularity conditions on the prior distribution. Moreover, the RD function admits closed-form expressions in several cases of practical interest and can otherwise be computed numerically, e.g., by the Blahut--Arimoto algorithm~\cite{Blahut1972IT,Arimoto1972IT}, or relaxed through the Shannon lower bound (SLB), which is asymptotically tight for broad classes of sources in the low-distortion regime~\cite{Zamir-Tightness,7556344-Koch}. Classical Cram\'er--Rao-type bounds are local Fisher-information bounds and can become tight in regular high-information or low-noise regimes, but such behavior is not a universal high-noise guarantee. In contrast, we show that the proposed RDB converges exactly to the zero-rate Bayes risk in the high-sensing-noise limit and is also first-order tight for a Gaussian test channel in the low-noise limit.

The RD perspective has also been applied to Gaussian ISAC in~\cite{liu2023deterministic}. That work interprets wireless sensing as a non-cooperative joint source--channel coding problem: the target is viewed as a passive information source, the parameter-to-sensing-channel mapping plays the role of a source-to-channel mapping, and the sensing receiver observes the corresponding channel output. A key distinction from classical cooperative source--channel coding is that the physical target cannot jointly encode a long sequence of independent parameter realizations into a block code; it can only induce a letter-wise physical mapping for each realization. This viewpoint explains why a general RD converse for sensing need not be operationally achievable in the same way as a classical separation theorem and why sensing tightness should be interpreted relative to the non-cooperative, per-coherence-block physical process~\cite{liu2023deterministic}.

Our result complements this viewpoint but strengthens the waveform conditioning. In~\cite{liu2023deterministic}, the RD function is applied to sensing information averaged over the random waveform. The converse developed here retains the sensing mutual information induced by each waveform realization and applies the distortion--rate function before averaging. Specifically, if
\begin{equation}
I_{\rm s}(\bfx)
\triangleq
I(\bfA;\bfY_{\rms}\mid\bfX=\bfx),
\end{equation}
our sensing bound contains
\begin{equation}
\bbE_{\bfX}
\left[
D_{\bfA,d}\!\left(I_{\rm s}(\bfX)\right)
\right],
\end{equation}
rather than $D_{\bfA,d}(\bbE_{\bfX}[I_{\rm s}(\bfX)])$. Since the distortion--rate function is convex in its information argument, Jensen's inequality shows that the former is no smaller than the latter, and it can be strictly larger when the waveform-dependent sensing information is nonconstant. In addition, our model permits the sensing state to be random conditioned on the parameter of interest, which naturally accommodates nuisance variables and mixture observation models.

Motivated by these observations, we develop a waveform-conditioned RDB converse that retains the dependence of sensing information on the realized transmitted signal. This formulation supports tractable relaxations and reveals how signal randomness and spatial energy allocation affect the sensing–communication tradeoff. Although problem-specific bounds may be tighter for particular sensing tasks, the proposed framework offers broad applicability and modularity.

\subsection{Contributions and Outline}
The main contributions are summarized as follows.
\begin{itemize}
    \item In Section~\ref{sec:general_isac}, we introduce a block-memoryless ISAC model with coherence time $T$ that allows continuous, discrete, and hybrid sensing parameters. The communication code spans many coherence blocks, while sensing is performed separately in each block. The sensing state is described by a conditional distribution given the parameter of interest, thereby including deterministic sensing maps as a special case and allowing latent nuisance variables.

    \item In Section~\ref{sec:general_isac}, we also state a general conditional RDB in Theorem~\ref{thm:Rate_Distortion_Bound} and specialize it in Corollary~\ref{cor:waveform_conditioned_RDB} to obtain the waveform-conditioned ISAC converse. The resulting outer region in Theorem~\ref{thm:generic_converse} applies the distortion--rate function to the sensing information associated with each realized waveform before averaging. By convexity, this form is no weaker than applying the distortion--rate function after averaging the sensing information, and it can be strictly tighter when that information varies across waveform realizations.

    \item In Section~\ref{sec:generic_converse}, we characterize two tightness regimes and relate the converse to the BCRB. The RDB is exact in the high-sensing-noise limit; under scalar squared-error loss, a conditional version of Stam's inequality shows that its entropy-power specialization is no looser than the BCRB; and a Gaussian test channel establishes first-order tightness at low sensing noise.

    \item In Section~\ref{sec:occupancy_detection}, we specialize the framework to binary occupancy detection under Hamming loss and a nuisance-induced mixture observation model. We derive waveform-dependent and covariance-based RDB converses, compare them with a task-specific total-variation bound, and evaluate Gaussian and semi-unitary achievable schemes with the optimal Bayesian detector.

    \item In Section~\ref{sec:Nakagami}, we study multiple-input multiple-output (MIMO) target-response estimation with i.i.d. Nakagami entries under MSE. We evaluate the Bayesian Fisher information, identify severe-fading regimes in which the BCRB yields only the trivial zero lower bound, derive an entropy-power RDB, and obtain covariance-based outer and achievable regions for Gaussian and semi-unitary signaling.
\end{itemize}

\subsection{Notation}
Deterministic scalars are denoted by lowercase letters, and random variables by uppercase letters. Bold letters represent vectors and matrices, and calligraphic letters denote alphabets and feasible sets. For a matrix $\mathbf A$, the notation $\mathbf A^{\sfT}$, $\mathbf A^{\dagger}$, $\mathbf A^{-1}$, $\det(\mathbf A)$, and $\operatorname{tr}(\mathbf A)$ denote its transpose, Hermitian transpose, inverse, determinant, and trace, respectively. For $\mathbf A\in\bbC^{n\times m}$, $\operatorname{vec}(\mathbf A)=[\mathbf A_1^{\sfT},\ldots,\mathbf A_m^{\sfT}]^{\sfT}$, where $\mathbf A_i$ is the $i$th column of $\mathbf A$. We denote the $k$-dimensional identity matrix by $\mathbf I_k$, the all-zero vector by $\mathbf0$, the semidefinite ordering by $\succeq$, and the Kronecker product by $\otimes$. The distribution of a random variable $\mathbf X$ is denoted by $P_{\mathbf X}$, and the closure of a convex hull by $\overline{\operatorname{conv}}$. Unless stated otherwise, logarithms, entropies, and mutual informations are measured in bits; $\ln(\cdot)$ denotes the natural logarithm.

\section{General ISAC Model and Main Converse}
\label{sec:general_isac}

\begin{figure}[t]
\centering
\resizebox{\linewidth}{!}{
\begin{tikzpicture}[
  block/.style = {rectangle, draw, minimum width=2cm, minimum height=1cm, align=center},
  arrow/.style = {-{Stealth}, thick}
]
  \node[block] (Tx) {ISAC\\Transmitter};
  \node[block, left=2.8cm of Tx] (Enc) {Encoder};
  \node[block, minimum width=1.2cm, above right=0cm and 2cm of Tx] (Wc) {$\bfH_{\rmc}$};
  \node[block, minimum width=1.2cm, below right=0cm and 2cm of Tx] (Ws) {$\bfH_{\rms}$};
  \node[block, minimum width=3cm, right=3.9cm of Wc] (CommRx) {Communication\\Receiver};
  \node[block, minimum width=3cm, right=3.9cm of Ws] (SensingRx) {Sensing\\Receiver};
  \coordinate (E1) at ($(Tx.north east)!0.333!(Tx.south east)$);
  \coordinate (E2) at ($(Tx.north east)!0.666!(Tx.south east)$);
  \coordinate (OutC) at ($(CommRx.east)+(1cm,0)$);
  \coordinate (OutS) at ($(SensingRx.east)+(1cm,0)$);
  \coordinate (Xmid) at ($(Enc.east)!0.5!(Tx.west)$);
  \coordinate (TapBelow) at ($(SensingRx.south)+(0,-0.5cm)$);
  \coordinate (TapAbove) at ($(CommRx.north)+(0,+0.5cm)$);
  \coordinate (TapAboveHc) at (Wc.north);
  \draw[arrow] (Enc.east) -- (Tx.west);
  \node[above] at (Xmid) {$\bfX_1,\bfX_2,\ldots,\bfX_T$};
  \draw[arrow] (Xmid) -- ++(0,-2.2cm) -| (TapBelow) -- (SensingRx.south);
  \draw[arrow] (TapAboveHc) -- ++(0,+1cm) -| (TapAbove) -- (CommRx.north);
  \draw[arrow] (E1) -| ++(0.75cm,0) |- (Wc.west);
  \draw[arrow] (E2) -| ++(0.75cm,0) |- (Ws.west);
  \draw[arrow] (Wc.east) -- (CommRx.west);
  \node[above] at ($(Wc)!0.42!(CommRx)$) {$\bfY_{{\rm c},1},\ldots,\bfY_{{\rm c},T}$};
  \draw[arrow] (Ws.east) -- (SensingRx.west);
  \node[above] at ($(Ws)!0.42!(SensingRx)$) {$\bfY_{{\rm s},1},\ldots,\bfY_{{\rm s},T}$};
  \draw[arrow] (SensingRx) -- (OutS);
  \draw[arrow] (CommRx) -- (OutC);
\end{tikzpicture}}
\caption{General block-memoryless ISAC model. Communication coding spans many coherence blocks of length $T$, whereas sensing is performed separately within each block.}
\label{fig:general_isac}
\end{figure}

\subsection{System Model}
We consider the block-memoryless ISAC model in Fig.~\ref{fig:general_isac}. Each coherence block contains $T$ channel uses. Across blocks, the channel tuples are independent copies of the same statistical model. A communication code spans $n$ coherence blocks, whereas the sensing receiver produces one estimate of the sensing parameter in each block. Thus, communication benefits from coding across blocks, while sensing remains a per-coherence-block inference task.

\begin{definition}[Block-memoryless ISAC channel with communication CSIR]
\label{def:dm_isac}
Let $\cX,\cY_{\rmc},\cY_{\rms},\cH_{\rmc},\cH_{\rms}$, and $\cA$ be measurable alphabets. A block-memoryless ISAC channel with coherence time $T$ is specified by the sensing parameter $\bfA\in\cA$ with prior $P_{\bfA}$, the channel states $(\bfH_{\rmc},\bfH_{\rms})$, the input block
\begin{equation}
\bfX=(\bfX_1,\ldots,\bfX_T)\in\cX^T,
\end{equation}
and the output blocks
\begin{equation}
\bfY_{\rmc}=(\bfY_{{\rm c},1},\ldots,\bfY_{{\rm c},T}),\qquad
\bfY_{\rms}=(\bfY_{{\rm s},1},\ldots,\bfY_{{\rm s},T}),
\end{equation}
whose conditional law factors as
\begin{equation}
P_{\bfA,\bfH_{\rmc},\bfH_{\rms},\bfY_{\rmc},\bfY_{\rms}\mid\bfX}
=
P_{\bfA}
P_{\bfH_{\rmc},\bfH_{\rms}\mid\bfA}
P_{\bfY_{\rmc},\bfY_{\rms}\mid\bfX,\bfH_{\rmc},\bfH_{\rms}}.
\label{eq:channelpdf}
\end{equation}
The channel is memoryless across coherence blocks. No independence assumption is imposed across the $T$ channel uses within a coherence block; any temporal dependence is included in the block transition law in~\eqref{eq:channelpdf}. The communication receiver knows $\bfH_{\rmc}$, and the sensing receiver knows the transmitted waveform $\bfX$, as in a monostatic architecture or any setting in which the sensing receiver has access to the transmitted block.
\end{definition}

The factorization in~\eqref{eq:channelpdf} implies
\begin{equation}
P_{\bfA\mid\bfX}=P_{\bfA},
\qquad\text{i.e.,}\qquad
\bfA\independent\bfX.
\label{eq:A_independent_X}
\end{equation}
This corresponds to an open-loop sensing waveform that does not depend on the unknown parameter to be estimated. Importantly,~\eqref{eq:A_independent_X} does not imply that $\bfX$ is irrelevant to sensing: the conditional sensing law $P_{\bfY_{\rms}\mid\bfA,\bfX}$, and hence the information available about $\bfA$, can depend strongly on the realized waveform.

The conditional law $P_{\bfH_{\rmc},\bfH_{\rms}\mid\bfA}$ allows the sensing state to remain random after conditioning on $\bfA$. This includes the deterministic model $\bfH_{\rms}=\bfg(\bfA)$ as the special case in which $P_{\bfH_{\rms}\mid\bfA=\bfa}$ is a point mass at $\bfg(\bfa)$, while also permitting latent nuisance parameters that affect the sensing state but are not themselves part of the estimation objective.

\begin{definition}[Sensing distortion]
\label{def:RDcode}
Fix a distortion function $d:\cA\times\hat{\cA}\to[0,\infty)$. For an admissible input law $P_{\bfX}$, the sensing receiver uses a measurable estimator
\begin{equation}
\hat{\bfA}=\psi_{\rm s}(\bfY_{\rms},\bfX),
\end{equation}
and the minimum achievable per-block sensing distortion is
\begin{equation}
D^{\star}(P_{\bfX})
\triangleq
\inf_{\psi_{\rm s}}
\bbE\left[d(\bfA,\psi_{\rm s}(\bfY_{\rms},\bfX))\right].
\label{eq:egk_block_distortion}
\end{equation}
A distortion level $D$ is feasible under $P_{\bfX}$ whenever $D\ge D^{\star}(P_{\bfX})$.
\end{definition}

\begin{definition}[Achievable communication rate]
\label{def:rate}
For an admissible input law $P_{\bfX}$, coding over many independent coherence blocks yields
\begin{equation}
R\le\frac{1}{T}I(\bfX;\bfY_{\rmc}\mid\bfH_{\rmc}).
\label{eq:ergodic_rate_achievable}
\end{equation}
\end{definition}

\begin{definition}[ISAC capacity--distortion region]
\label{def:capacity}
Let $\cF\subseteq\cP(\cX^T)$ denote the set of admissible input laws. For example, $\cF$ may impose an average block-power constraint $\operatorname{tr}(\bbE[\bfX\bfX^\dagger])\le P$ for a prescribed block-energy budget $P$. The ISAC capacity--distortion region is
{%\small
\begin{align}
\cC
\triangleq
\overline{\operatorname{conv}}
\bigcup_{P_{\bfX}\in\cF}
\Big\{(R,D):\;&
R\le\frac{1}{T}I(\bfX;\bfY_{\rmc}\mid\bfH_{\rmc}),
\nonumber\\[-1mm]
& D\ge D^{\star}(P_{\bfX})
\Big\}.
\label{eq:capacity_distortion_region}
\end{align}
}
\end{definition}

\subsection{Conditional Rate--Distortion Converse}
\label{sec:rate-distortion}
For a source $\sfV\in\cV$ and distortion measure $d$, the rate--distortion function is~\cite{Cover:2006}
\begin{equation}
R_{\sfV,d}(D)
\triangleq
\inf_{P_{\hat{\sfV}\mid\sfV}:\,\bbE[d(\sfV,\hat{\sfV})]\le D}
I(\sfV;\hat{\sfV}).
\label{eq:alphaRDF}
\end{equation}
Equivalently, define the distortion--rate function
\begin{equation}
D_{\sfV,d}(r)
\triangleq 
\inf_{P_{\hat{\sfV}\mid\sfV}:\,I(\sfV;\hat{\sfV})\le r}
\bbE[d(\sfV,\hat{\sfV})]
= 
\inf\{D:R_{\sfV,d}(D)\le r\}.
\label{eq:distortion_rate_function}
\end{equation}
The function $D_{\sfV,d}(r)$ is non-increasing and convex in $r$, and its zero-rate value is
\begin{equation}
D_{\sfV,d}(0^+)=\inf_c\bbE[d(\sfV,c)].
\label{eq:inverseinZero}
\end{equation}

For each realization $\sfU=\sfu$, let $D_{\sfV\mid\sfU=\sfu,d}(r)$ denote the distortion--rate function associated with the conditional source law $P_{\sfV\mid\sfU=\sfu}$:
{%\small
\begin{align}
D_{\sfV\mid\sfU=\sfu,d}(r)
\triangleq \hspace*{-.5cm}
\inf_{P_{\hat{\sfV}\mid\sfV,\sfU=\sfu}:\,
I(\sfV;\hat{\sfV}\mid\sfU=\sfu)\le r}
\bbE[d(\sfV,\hat{\sfV})\mid\sfU=\sfu].
\label{eq:conditional_distortion_rate}
\end{align}
}

\begin{theorem}[General conditional RDB]
\label{thm:Rate_Distortion_Bound}
Let $\sfV$, $\sfW$, and $\sfU$ be arbitrarily jointly distributed random variables. For any measurable estimator $\hat{\sfV}=g(\sfW,\sfU)$,
\begin{align}
\bbE[d(\sfV,g(\sfW,\sfU))]
\ge
\bbE_{\sfU}
\left[
D_{\sfV\mid\sfU=\sfu,d}
\left(I(\sfV;\sfW\mid\sfU=\sfu)\right)
\right].
\label{eq:general_conditional_RDB}
\end{align}
\end{theorem}

\begin{proof}
Fix $\sfU=\sfu$ and define $\hat{\sfV}_{\sfu}=g(\sfW,\sfu)$. Conditioned on $\sfU=\sfu$, the Markov chain $\sfV\longrightarrow\sfW\longrightarrow\hat{\sfV}_{\sfu}$ and the data-processing inequality give
\begin{equation}
I(\sfV;\hat{\sfV}_{\sfu}\mid\sfU=\sfu)
\le
I(\sfV;\sfW\mid\sfU=\sfu).
\end{equation}
The induced reconstruction is therefore feasible in~\eqref{eq:conditional_distortion_rate} at information budget $I(\sfV;\sfW\mid\sfU=\sfu)$. Applying the definition conditionally and then averaging over $P_{\sfU}$ proves~\eqref{eq:general_conditional_RDB}.
\end{proof}

\begin{corollary}[Waveform-conditioned ISAC RDB]
\label{cor:waveform_conditioned_RDB}
For the ISAC model of Definition~\ref{def:dm_isac}, any admissible $P_{\bfX}\in\cF$ and measurable estimator $\hat{\bfA}=\psi_{\rm s}(\bfY_{\rms},\bfX)$ satisfy
\begin{equation}
\bbE[d(\bfA,\hat{\bfA})]
\ge
\bbE_{\bfX}
\left[
D_{\bfA,d}\!\left(
I(\bfA;\bfY_{\rms}\mid\bfX=\bfx)
\right)
\right].
\label{eq:waveform_conditioned_RDB}
\end{equation}
Consequently,
\begin{equation}
D^{\star}(P_{\bfX})
\ge
\bbE_{\bfX}
\left[
D_{\bfA,d}\!\left(
I(\bfA;\bfY_{\rms}\mid\bfX=\bfx)
\right)
\right].
\label{eq:waveform_conditioned_RDB_opt}
\end{equation}
\end{corollary}

\begin{proof}
Apply Theorem~\ref{thm:Rate_Distortion_Bound} with $(\sfV,\sfW,\sfU)=(\bfA,\bfY_{\rms},\bfX)$. By~\eqref{eq:A_independent_X}, $P_{\bfA\mid\bfX=\bfx}=P_{\bfA}$, so the conditional distortion--rate function reduces to $D_{\bfA,d}$ for $P_{\bfX}$-almost every $\bfx$.
\end{proof}

\begin{rem}[Relation to an averaged RDB]
\label{rem:conditional_tightness}
Since $D_{\bfA,d}(r)$ is convex, we have
\begin{align}
&\bbE_{\bfX}\left[
D_{\bfA,d}\!\left(I(\bfA;\bfY_{\rms}\mid\bfX=\bfx)\right)
\right]
%\nonumber\\&\qquad
\ge
D_{\bfA,d}\!\left(I(\bfA;\bfY_{\rms}\mid\bfX)\right).
\label{eq:jensen_conditional_RDB}
\end{align}
Thus, retaining the waveform realization inside the distortion--rate function before averaging yields a sensing converse that is no weaker than the corresponding waveform-averaged RDB. The inequality can be strict when $I(\bfA;\bfY_{\rms}\mid\bfX=\bfx)$ varies with $\bfx$ and $D_{\bfA,d}(r)$ is strictly convex over the relevant range.
\end{rem}

\subsection{ISAC Outer Region}
\label{sec:gen_conve_subsection}
For convenience, define
\begin{equation}
\underline D_{\rm RDB}(P_{\bfX})
\triangleq
\bbE_{\bfX}\left[
D_{\bfA,d}\!\left(I(\bfA;\bfY_{\rms}\mid\bfX=\bfx)\right)
\right].
\label{eq:generic_sensing_RDB}
\end{equation}

\begin{theorem}[ISAC region outer bound]
\label{thm:generic_converse}
The ISAC capacity--distortion region satisfies $\cC\subseteq\cR$, where
\begin{align}
\cR
\triangleq
\overline{\operatorname{conv}}
\bigcup_{P_{\bfX}\in\cF}
\Bigg\{(R,D):\quad
&R\le\frac{1}{T}I(\bfX;\bfY_{\rmc}\mid\bfH_{\rmc}),
\nonumber\\[-1mm]
&D\ge\underline D_{\rm RDB}(P_{\bfX})
\Bigg\}.
\label{eq:generic_outer_region}
\end{align}
\end{theorem}

\begin{proof}
For every $P_{\bfX}\in\cF$, Corollary~\ref{cor:waveform_conditioned_RDB} lower-bounds $D^{\star}(P_{\bfX})$, while~\eqref{eq:ergodic_rate_achievable} upper-bounds the corresponding communication rate. Taking the union over admissible input laws and its convex closure proves the result.
\end{proof}

The converse in Theorem~\ref{thm:generic_converse} requires no regularity conditions on the joint distribution. This should be contrasted with bounds such as the BCRB, which require smoothness and appropriate behavior at the boundary of the support. The occupancy-detection and Nakagami examples below illustrate sensing settings in which the RDB remains applicable while the conventional BCRB is not directly available or informative.

\section{Computable RDB Specializations and Tightness}
\label{sec:generic_converse}
Theorem~\ref{thm:generic_converse} is stated in terms of the distortion--rate function $D_{\bfA,d}(r)$. Although this function is known in closed form for several source--distortion pairs, it may be difficult to evaluate in general. Classical rate--distortion lower bounds therefore provide useful computable relaxations. We first collect the squared-error specializations required below and then discuss tightness and the relation to the BCRB.

\subsection{Useful Distortion--Rate Specializations}
\label{sec:RDB_specializations}

\begin{lemma}[Shannon lower bound for squared error]
\label{lemma:SLB_continuous}
Let $\sfV$ be a continuous random vector with finite differential entropy. If $\sfV\in\bbR^n$ and $d(\sfv,\hat{\sfv})=\|\sfv-\hat{\sfv}\|_2^2$, then
\begin{equation}
D_{\sfV,d}(r)
\ge
\frac{n}{2\pi\rme}
2^{\frac{2}{n}(h(\sfV)-r)}.
\label{eq:SLB_inv_continuous}
\end{equation}
If $\sfV\in\bbC^n$ and $d(\sfv,\hat{\sfv})=\|\sfv-\hat{\sfv}\|_2^2$, then
\begin{equation}
D_{\sfV,d}(r)
\ge
\frac{n}{\pi\rme}
2^{\frac{1}{n}(h(\sfV)-r)}.
\label{eq:SLB_inv_complex}
\end{equation}
\end{lemma}
These are direct specializations of the classical Shannon lower bound~\cite{Cover:2006,riegler2018rate}. Broader continuous and discrete forms are collected in Appendix~\ref{app:general_SLB}.

The same construction produces a computable but potentially looser ISAC outer region whenever a Shannon-type lower bound
\begin{equation}
\underline D_{\bfA,d}^{\rm SLB}(r)\le D_{\bfA,d}(r)
\label{eq:generic_DSLB_definition}
\end{equation}
is available. Define
\begin{equation}
\underline D_{\rm SLB}(P_{\bfX})
\triangleq
\bbE_{\bfX}\left[
\underline D_{\bfA,d}^{\rm SLB}\!\left(
I(\bfA;\bfY_{\rms}\mid\bfX=\bfx)
\right)
\right]
\label{eq:generic_sensing_SLB}
\end{equation}
and
\begin{align}
\cR_{\rm SLB}
\triangleq
\overline{\operatorname{conv}}
\bigcup_{P_{\bfX}\in\cF}
\Bigg\{(R,D):\quad
&R\le\frac{1}{T}I(\bfX;\bfY_{\rmc}\mid\bfH_{\rmc}),
\nonumber\\[-1mm]
&D\ge\underline D_{\rm SLB}(P_{\bfX})
\Bigg\}.
\label{eq:generic_SLB_outer_region}
\end{align}
Since~\eqref{eq:generic_DSLB_definition} implies $\underline D_{\rm RDB}(P_{\bfX})\ge\underline D_{\rm SLB}(P_{\bfX})$ for every admissible input law,
\begin{equation}
\cC\subseteq\cR\subseteq\cR_{\rm SLB}.
\label{eq:C_R_RSLB}
\end{equation}

\subsection{High-Sensing-Noise Tightness}
\label{sec:tightness_gen_converse}
Let the sensing channel be indexed by a noise parameter $\eta\ge0$, with observation $\bfY_{\rms}^{(\eta)}$, and assume
\begin{equation}
\lim_{\eta\to\infty}
I(\bfA;\bfY_{\rms}^{(\eta)}\mid\bfX=\bfx)=0
\quad\text{for $P_{\bfX}$-a.e. $\bfx$}.
\label{eq:high_noise_condition}
\end{equation}

\begin{prop}[Tightness in the high sensing-noise regime]
\label{prop:tightness_gen_Converse}
Suppose $D_{\bfA,d}(0^+)<\infty$. For any fixed $P_{\bfX}\in\cF$ satisfying~\eqref{eq:high_noise_condition},
\begin{align}
\lim_{\eta\to\infty}D_\eta^\star(P_{\bfX})
&=
\lim_{\eta\to\infty}
\bbE_{\bfX}\left[
D_{\bfA,d}\!\left(
I(\bfA;\bfY_{\rms}^{(\eta)}\mid\bfX=\bfx)
\right)
\right]
\nonumber\\
&=
D_{\bfA,d}(0^+)
=
\inf_{\bfc}\bbE[d(\bfA,\bfc)].
\label{eq:high_noise_tightness}
\end{align}
\end{prop}
\begin{proof}
See Appendix~\ref{app:high_noise_tightness}.
\end{proof}

\subsection{Comparison with the Bayesian Cram\'er--Rao Bound}
\label{sec:comparison_to_CRB}
Consider a continuous random vector $\sfV\in\bbR^n$ estimated from $\sfW$. From Theorem~\ref{thm:Rate_Distortion_Bound} with constant $\sfU$ and Lemma~\ref{lemma:SLB_continuous},
\begin{align}
\operatorname{MMSE}(\sfV\mid\sfW)
&\ge
D_{\sfV,d}(I(\sfV;\sfW))
\nonumber\\
&\ge
\frac{n}{2\pi\rme}
2^{\frac{2}{n}h(\sfV\mid\sfW)}
=
nN(\sfV\mid\sfW),
\label{eq:conditional_version_RDB}
\end{align}
where
\begin{equation}
N(\sfV\mid\sfW)
\triangleq
\frac{1}{2\pi\rme}
2^{\frac{2}{n}h(\sfV\mid\sfW)}.
\label{eq:conditional_entropy_power}
\end{equation}
Whenever the BCRB is well defined~\cite{miller1978modified},
\begin{equation}
\operatorname{MMSE}(\sfV\mid\sfW)
\ge
\operatorname{tr}(\bfJ^{-1}(\sfV\mid\sfW)).
\label{eq:conditional_version_BCRB}
\end{equation}

\begin{prop}[Scalar RDB is not looser than the BCRB]
\label{prop:salardominance}
For scalar $\sfV\in\bbR$,
\begin{align}
\operatorname{MMSE}(\sfV\mid\sfW)
&\ge D_{\sfV,d}(I(\sfV;\sfW))
\nonumber\\
&\ge N(\sfV\mid\sfW)
\ge J^{-1}(\sfV\mid\sfW).
\label{eq:RDB_BCRB_ordering}
\end{align}
\end{prop}
The last inequality follows from the conditional version of Stam's inequality,
\begin{equation}
N(\sfV\mid\sfW)\ge J^{-1}(\sfV\mid\sfW),
\label{eq:conditional_Stam}
\end{equation}
proved in Appendix~\ref{app:conditional_Stam}. The scalar argument does not establish an ordering between conditional entropy power and a matrix- or trace-valued BCRB in higher dimensions. We therefore make no vector dominance claim; establishing one, if possible under additional assumptions, would require a genuinely matrix-valued comparison inequality.

\subsection{Low-Sensing-Noise Behavior}
\label{sec:low_noise_tightness}
Consider
\begin{equation}
\sfW=\sfV+\sigma\sfZ,
\qquad
\sfZ\sim\cN({\bf0},\bfI_n),
\qquad
\sfZ\independent\sfV.
\label{eq:GaussChannel}
\end{equation}
Assume that $\sfV$ is absolutely continuous, has finite second moment and finite differential entropy, and satisfies the Gaussian-smoothing continuity condition $h(\sfV+\sigma\sfZ)\to h(\sfV)$ as $\sigma\to0$. Then
\begin{align}
\lim_{\sigma\to0}
\frac{D_{\rm SLB}(\sigma)}{\sigma^2}
&=
\lim_{\sigma\to0}
\frac{D_{\sfV,d}(I(\sfV;\sfW))}{\sigma^2}
\nonumber\\
&=
\lim_{\sigma\to0}
\frac{\operatorname{MMSE}(\sfV\mid\sfW)}{\sigma^2}
=n,
\label{eq:low_noise_joint_limit}
\end{align}
where
\begin{equation}
D_{\rm SLB}(\sigma)
\triangleq
\frac{n}{2\pi\rme}
2^{\frac{2}{n}(h(\sfV)-I(\sfV;\sfW))}.
\end{equation}
Thus, both the entropy-power specialization and the exact RDB are first-order tight for the Gaussian test channel. The proof is given in Appendix~\ref{app:low_noise_tightness}.

\section{Binary Occupancy Detection}
\label{sec:occupancy_detection}
We first specialize the general converse to a binary sensing task in which the sensing receiver decides whether a person is present in a monitored environment. This example highlights two features of the proposed framework: it applies directly to a discrete sensing parameter under Hamming loss, and it accommodates sensing channels that remain random after conditioning on the parameter of interest because of latent nuisance variables. In particular, under Hamming loss the sensing distortion is exactly the occupancy error probability, rather than a proxy for binary detection performance, such as the Kullback-Leibler divergence adopted as a sensing metric in~\cite{Fei24KL}.

\subsection{ISAC Model}
\label{sec:ISAC_model_occupancy}
Let
\begin{equation}
A\sim\operatorname{Bern}(p_A),
\end{equation}
where $A=0$ denotes an empty environment and $A=1$ an occupied environment. The communication and sensing observations are
\begin{equation}
\bfY_{\star}=\bfH_{\star}\bfX+\bfZ_{\star},
\qquad
\star\in\{\rmc,\rms\},
\label{eq:occ_gaussian_channels}
\end{equation}
where the entries of $\bfZ_{\star}$ are i.i.d. $\cC\cN(0,\sigma_{\star}^{2})$. The transmit block $\bfX\in\bbC^{M\times T}$ satisfies
\begin{equation}
\cF
=
\left\{
P_{\bfX}:
\operatorname{tr}\left(\bbE[\bfX\bfX^\dagger]\right)
\le MTP_0
\right\}.
\label{eq:APowerConstraint_occupancy}
\end{equation}

The sensing channel is modeled as a known static response plus a rank-one perturbation induced by occupancy,
\begin{equation}
\bfH_{\rms}
=
\bfH_0+A\alpha_{\rms} \, \bfu(\Theta)\bfv^\dagger(\Theta),
\label{eq:occ_sensing_model_los_rewrite}
\end{equation}
where $\bfH_0\in\bbC^{N_{\rms}\times M}$ is the static empty-environment response, e.g., the aggregate response produced by walls, ceiling, and other stationary reflectors,
\begin{equation}
\alpha_{\rms}\sim\cC\cN(0,\sigma_{\alpha_{\rms}}^2),
\end{equation}
is the scattering coefficient of the additional occupancy-induced reflector, e.g., a human body, and $\Theta$ is its unknown angle. The coefficient $\alpha_{\rms}$ and angle $\Theta$ are nuisance variables and are not part of the sensing objective. The unit-norm half-wavelength ULA steering vectors are
\begin{align}
\bfv(\Theta)
&=
\frac{1}{\sqrt M}
\left[1,\rme^{j\pi\sin\Theta},\ldots,\rme^{j\pi(M-1)\sin\Theta}\right]^{\sfT},
\label{eq:ula_v_occ_rewrite}\\
\bfu(\Theta)
&=
\frac{1}{\sqrt{N_{\rms}}}
\left[1,\rme^{j\pi\sin\Theta},\ldots,\rme^{j\pi(N_{\rms}-1)\sin\Theta}\right]^{\sfT}.
\label{eq:ula_u_occ_rewrite}
\end{align}
The sensing receiver knows $(\bfY_{\rms},\bfX,\bfH_0)$ but not $(\Theta,\alpha_{\rms})$.

For communication, we consider the rank-one far-field LoS channel
\begin{equation}
\bfH_{\rmc}
=
\alpha_{\rmc} \, \bfu_{\rmc}(\phi)\bfv_{\rmc}^\dagger(\varphi),
\qquad
\alpha_{\rmc}\sim\cC\cN(0,1),
\label{eq:los_comm_channel_occ_rewrite}
\end{equation}
where $\bfu_{\rmc}\in\bbC^{N_{\rmc}}$ and $\bfv_{\rmc}\in\bbC^M$ are deterministic unit-norm steering vectors. The communication directions $(\phi,\varphi)$ are known at the transmitter and receiver, whereas $\alpha_{\rmc}$ is known only at the communication receiver.

We use Hamming distortion,
\begin{equation}
d(A,\hat A)=\mathbbm1\{A\neq\hat A\},
\end{equation}
so that
\begin{equation}
D=\bbP[\hat A\neq A].
\end{equation}
In the remainder of this section, we specialize to the maximally uncertain binary prior,
\begin{equation}
P_A(0)=P_A(1)=\frac12.
\label{eq:uniform_occupancy_prior}
\end{equation}
This specialization is not required by the general RDB framework, but simplifies the expressions that follow.

\subsection{Waveform-Dependent Occupancy Converse}
\label{subsec:occupancy_waveform_converse}
Since $\bfH_0\bfX$ is known at the sensing receiver, define the clutter-canceled residual
\begin{equation}
\widetilde{\bfY}_{\rms}
\triangleq
\bfY_{\rms}-\bfH_0\bfX
=
A\alpha_{\rms} \, \bfu(\Theta)\bfv^\dagger(\Theta)\bfX+\bfZ_{\rms},
\label{eq:occupancy_residual}
\end{equation}
and let
\begin{equation}
\widetilde{\sfY}_{\rms}
\triangleq
\operatorname{vec}(\widetilde{\bfY}_{\rms})
\in\bbC^{N_{\rms}T}.
\end{equation}
For a fixed waveform $\bfX=\bfx$, denote the conditional densities under $A=0$ and $A=1$ by
\begin{equation}
q_a(\sfy\mid\bfx)
\triangleq
f_{\widetilde{\sfY}_{\rms}\mid A=a,\bfX=\bfx}(\sfy),
\qquad a\in\{0,1\}.
\end{equation}
Under $A=0$,
\begin{equation}
q_0(\cdot\mid\bfx)
=
\cC\cN({\bf0},\sigma_{\rms}^2\bfI).
\label{eq:q0_def}
\end{equation}
Under $A=1$, conditioning on $\Theta=\theta$ and marginalizing $\alpha_{\rms}$ gives
\begin{equation}
q_{1,\theta}(\cdot\mid\bfx)
=
\cC\cN({\bf0},\boldsymbol{\Sigma}_{1,\theta}(\bfx)),
\label{eq:q1_theta_gauss_rewrite}
\end{equation}
where
\begin{align}
\boldsymbol{\Sigma}_{1,\theta}(\bfx)
&=
\sigma_{\rms}^2\bfI
+
{\sigma_{\alpha_{\rms}}}^2\bfg_\theta(\bfx)\bfg_\theta^\dagger(\bfx),\\
\bfg_\theta(\bfx)
&\triangleq
\operatorname{vec}\!\left(\bfu(\theta)\bfv^\dagger(\theta)\bfx\right)
=
\left(\bfx^{\sfT}\bfv^*(\theta)\right)\otimes\bfu(\theta).
\label{eq:gtheta_occ}
\end{align}
Since $\Theta$ is unknown, the occupied-state likelihood is the mixture
\begin{equation}
q_1(\sfy\mid\bfx)
=
\bbE_\Theta[q_{1,\Theta}(\sfy\mid\bfx)].
\label{eq:q1_def}
\end{equation}

For two densities $p_0$ and $p_1$, define their Bhattacharyya coefficient as
\begin{equation}
Z(p_0,p_1)
\triangleq
\int\sqrt{p_0(\sfy)p_1(\sfy)}\,\rmd\sfy.
\label{eq:Bhattacharyya_coefficient}
\end{equation}
For proper complex zero-mean Gaussian densities $p_i=\cC\cN({\bf0},\boldsymbol{\Sigma}_i)$,
\begin{equation}
Z(p_0,p_1)
=
\frac{
\det(\boldsymbol{\Sigma}_0)^{1/2}
\det(\boldsymbol{\Sigma}_1)^{1/2}
}{
\det\!\left((\boldsymbol{\Sigma}_0+\boldsymbol{\Sigma}_1)/2\right)
}.
\label{eq:bhatt_complex_general_rewrite}
\end{equation}
A derivation is provided in Appendix~\ref{app:occupancy_bhattacharyya}.

Applying~\eqref{eq:bhatt_complex_general_rewrite} to $q_0$ and $q_{1,\theta}$ and using the matrix determinant lemma gives
\begin{equation}
Z(q_0(\cdot\mid\bfx),q_{1,\theta}(\cdot\mid\bfx))
=
z(t_\theta(\bfx)),
\label{eq:Z_theta_closed_form_rewrite}
\end{equation}
where
\begin{equation}
z(t)\triangleq\frac{\sqrt{1+t}}{1+t/2},\qquad t\ge0,
\label{eq:z_occ}
\end{equation}
and
\begin{align}
t_\theta(\bfx)
&\triangleq
\frac{\sigma_{\alpha_{\rms}}^2}{\sigma_{\rms}^2}
\|\bfx^\dagger\bfv(\theta)\|^2
%\nonumber\\&
=
\frac{\sigma_{\alpha_{\rms}}^2}{\sigma_{\rms}^2}
\bfv^\dagger(\theta)\bfx\bfx^\dagger\bfv(\theta).
\label{eq:t_theta_occ_rewrite}
\end{align}
The scalar $t_\theta(\bfx)$ is the effective sensing SNR associated with the energy radiated by waveform $\bfx$ toward direction $\theta$.

Since $q_1$ is a mixture over $\Theta$ and the square-root function is concave,
\begin{equation}
Z(q_0(\cdot\mid\bfx),q_1(\cdot\mid\bfx))
\ge
\bbE_\Theta[z(t_\Theta(\bfx))].
\label{eq:Z_mix_lower_rewrite}
\end{equation}
Define
\begin{equation}
Z_{\rm L}(\bfx)
\triangleq
\bbE_\Theta[z(t_\Theta(\bfx))].
\label{eq:ZL_occ}
\end{equation}
For equiprobable binary hypotheses,
\begin{align}
I(A;\bfY_{\rms}\mid\bfX=\bfx)
&\le
\operatorname{TV}(q_0(\cdot\mid\bfx),q_1(\cdot\mid\bfx))
\nonumber\\
&\le
\sqrt{1-Z_{\rm L}(\bfx)^2}.
\label{eq:MI_ub_from_BC_rewrite}
\end{align}
where the second inequality follows from the standard relation between total variation and Bhattacharyya overlap~\cite{Tsybakov2009Nonparametric}.

For the equiprobable Bernoulli source in~\eqref{eq:uniform_occupancy_prior} under Hamming distortion, the distortion--rate function is
\begin{equation}
D_{A,d}(r)=H_2^{-1}(1-r),
\qquad 0\le r\le1,
\label{eq:binary_distortion_rate_uniform}
\end{equation}
where $H_2^{-1}(\cdot)$ is restricted to $[0,1/2]$.

\begin{prop}[RDB lower bound on occupancy error]
\label{prop:RDBoccupancy}
Every sensing detector satisfies
\begin{equation}
\bbP[\hat A\neq A]
\ge
\bbE_{\bfX}
\left[
H_2^{-1}\!\left(1-\sqrt{1-Z_{\rm L}(\bfX)^2}\right)
\right].
\label{eq:Pe_outer_occupancy}
\end{equation}
\end{prop}

\begin{proof}
Combine~\eqref{eq:binary_distortion_rate_uniform}, Corollary~\ref{cor:waveform_conditioned_RDB}, and~\eqref{eq:MI_ub_from_BC_rewrite}.
\end{proof}

For comparison, the optimal error probability for equiprobable binary hypotheses satisfies
\begin{equation}
P_e^\star(\bfx)
=
\frac12\left[1-\operatorname{TV}(q_0(\cdot\mid\bfx),q_1(\cdot\mid\bfx))\right].
\end{equation}
Hence,
\begin{equation}
\bbP[\hat A\neq A]
\ge
\frac12\bbE_{\bfX}
\left[1-\sqrt{1-Z_{\rm L}(\bfX)^2}\right].
\label{eq:TV_bound_waveform}
\end{equation}
Unlike the RDB,~\eqref{eq:TV_bound_waveform} exploits the binary hypothesis-testing structure directly and therefore serves as a task-specific benchmark for interpreting the RDB in this binary setting.

\subsection{Covariance-Based ISAC Converse}
\label{subsec:outer_bound_bhatt_complex_rewrite}
Define the average transmit covariance
\begin{equation}
\bfQ\triangleq\bbE[\bfX\bfX^\dagger]
\end{equation}
and feasible set
\begin{equation}
\cQ
\triangleq
\{\bfQ\succeq{\bf0}:\operatorname{tr}(\bfQ)\le MTP_0\}.
\label{eq:Qset_occ}
\end{equation}
The function $z(t)$ in~\eqref{eq:z_occ}
is not globally convex. We therefore introduce the convex minorant
\begin{equation}
\underline z(t)
\triangleq
\begin{cases}
1+a_0t,&0\le t\le t_0,\\
z(t),&t\ge t_0,
\end{cases}
\label{eq:def_und_z}
\end{equation}
where
\begin{equation}
z'(t_0)=\frac{z(t_0)-1}{t_0},
\label{eq:t0_occ_definition}
\end{equation}
and $a_0=z'(t_0)$. Numerically,
\begin{equation}
t_0\simeq2.3830,
\qquad
a_0\simeq-0.0674.
\label{eq:t0_a0_occ_numeric}
\end{equation}
The values in~\eqref{eq:t0_a0_occ_numeric} follow from the tangency condition~\eqref{eq:t0_occ_definition}. The function $\underline z(t)$ is convex on $[0,\infty)$ and satisfies $\underline z(t)\le z(t)$; the derivation is given in Appendix~\ref{app:occupancy_covariance}.

For $\bfQ\in\cQ$, define
\begin{equation}
\zeta(\bfQ)
\triangleq
\bbE_\Theta
\left[
\underline z\!\left(
\frac{\sigma_{\alpha_{\rms}}^2}{\sigma_{\rms}^2}
\bfv^\dagger(\Theta)\bfQ\bfv(\Theta)
\right)
\right]
\label{eq:zeta_Q_occ}
\end{equation}
and
\begin{align}
\Phi_{\rm RDB}(\bfQ)
&\triangleq
H_2^{-1}\!\left(1-\sqrt{1-\zeta(\bfQ)^2}\right),
\label{eq:Phi_RDB_occ}\\
\Phi_{\rm TV}(\bfQ)
&\triangleq
\frac12\left(1-\sqrt{1-\zeta(\bfQ)^2}\right).
\label{eq:Phi_TV_occ}
\end{align}

For the rank-one communication channel~\eqref{eq:los_comm_channel_occ_rewrite},
\begin{equation}
R
\le
\overline R(\bfQ)
\triangleq
\frac{1}{\ln2}
\exp\!\left(
\frac{\sigma_{\rmc}^2T}{\bfv_{\rmc}^\dagger\bfQ\bfv_{\rmc}}
\right)
E_1\!\left(
\frac{\sigma_{\rmc}^2T}{\bfv_{\rmc}^\dagger\bfQ\bfv_{\rmc}}
\right),
\label{eq:Rbar_occ}
\end{equation}
where $E_1(\cdot)$ is the first-order exponential integral~\cite{alouini1999capacity}.

\begin{theorem}[Covariance-based occupancy converse]
\label{thm:converse_occupancy}
For the model of Section~\ref{sec:ISAC_model_occupancy},
\begin{align}
\cC
\subseteq
\overline{\operatorname{conv}}
\bigcup_{\bfQ\in\cQ}
\left\{
(R,D):
R\le\overline R(\bfQ),\;
D\ge\Phi_{\rm RDB}(\bfQ)
\right\}.
\label{eq:RDB_occupancy_relaxed}
\end{align}
Replacing $\Phi_{\rm RDB}(\bfQ)$ by $\Phi_{\rm TV}(\bfQ)$ gives the corresponding TV-based outer region.
\end{theorem}

\begin{proof}
The covariance reduction follows from the monotonicity and convexity of the outer maps and the convexity of $\underline z(t)$; a detailed derivation is given in Appendix~\ref{app:occupancy_covariance}. The communication expression follows by Gaussian maximization of the output entropy for fixed covariance and evaluation over $|\alpha_{\rmc}|^2\sim\operatorname{Exp}(1)$.
\end{proof}

For numerical evaluation, the RDB Pareto boundary is traced by
\begin{equation}
\bfQ_\lambda
\in
\argmin_{\bfQ\in\cQ}
\left\{
\Phi_{\rm RDB}(\bfQ)-\lambda\overline R(\bfQ)
\right\},
\qquad\lambda\ge0,
\label{eq:optimization_Qlambda_occupancy_relaxed}
\end{equation}
with the analogous optimization obtained by replacing $\Phi_{\rm RDB}$ by $\Phi_{\rm TV}$.

\subsection{Achievable Signaling and Detection}
\label{sec:occ_achievability_rewrite}
To assess tightness, consider covariances that allocate a fraction $\beta\in[0,1]$ of the block energy to the communication steering direction and distribute the remaining energy isotropically over its orthogonal complement:
{%\small
\begin{equation}
\bfQ_\beta
\triangleq
\beta MTP_0\bfv_{\rmc}\bfv_{\rmc}^\dagger
+
\frac{(1-\beta)MTP_0}{M-1}
(\bfI_M-\bfv_{\rmc}\bfv_{\rmc}^\dagger).
\label{eq:Qbeta_occ}
\end{equation}
}

For Gaussian signaling, the columns are independent and distributed as
\begin{equation}
\bfX_t\sim\cC\cN\!\left({\bf0},\frac{\bfQ_\beta}{T}\right),
\qquad t=1,\ldots,T.
\label{eq:Gaussian_occ_signalling}
\end{equation}
The coherent ergodic communication rate is
\begin{align}
R_{\rm G}(\beta)
&=
\bbE\left[
\log\!\left(1+\frac{\beta MP_0}{\sigma_{\rmc}^2}|\alpha_{\rmc}|^2\right)
\right]
\nonumber\\
&=
\frac{1}{\ln2}
\exp\!\left(\frac{\sigma_{\rmc}^2}{\beta MP_0}\right)
E_1\!\left(\frac{\sigma_{\rmc}^2}{\beta MP_0}\right).
\label{eq:R_G_occ_rewrite}
\end{align}

For a semi-unitary benchmark, let $\bfs\in\bbC^{1\times T}$ be isotropically distributed on the unit sphere, let $\bfS_\perp\in\bbC^{(M-1)\times T}$ satisfy $\bfS_\perp\bfS_\perp^\dagger=\bfI_{M-1}$ and be independent of $\bfs$, and let $\bfB_\perp$ have orthonormal columns spanning $\operatorname{null}(\bfv_{\rmc}^\dagger)$. We transmit
\begin{align}
\bfX_{\rm U}
&=
\sqrt{\beta MTP_0}\,\bfv_{\rmc}\bfs,\\
\bfX_{{\rm U},\perp}
&=
\sqrt{\frac{(1-\beta)MTP_0}{M-1}}\,\bfB_\perp\bfS_\perp,\\
\bfX&=\bfX_{\rm U}+\bfX_{{\rm U},\perp}.
\label{eq:X_U_occ_rewrite}
\end{align}
Both signaling constructions realize the covariance $\bbE[\bfX\bfX^\dagger]=\bfQ_\beta$; for the structured construction, the cross terms vanish because $\bfs$ is independent and zero mean. At high communication SNR, the semi-unitary ergodic rate is approximated by~\cite{ISAC_Caire}
\begin{align}
R_{\rm U}(\beta)
&\approx
\left(1-\frac{1}{2T}\right)
\left[
\log\!\left(\frac{\beta MP_0}{\sigma_{\rmc}^2}\right)
-\frac{\gamma_E}{\ln2}
\right]
\nonumber\\&+
\frac{1}{T}
\left[
\left(T-\frac12\right)\log\!\left(\frac{T}{\rme}\right)
+
\log\!\left(\frac{2\sqrt\pi}{\Gamma(T)}\right)
\right].
\label{eq:R_U_HSNR_occ_rewrite}
\end{align}

For either signaling strategy, the sensing receiver uses the optimal Bayesian detector. For fixed $\bfX=\bfx$, the angle-conditioned log-likelihood ratio is
\begin{align}
\ell(\theta;\sfy\mid\bfx)
\triangleq\;&
\ln\frac{q_{1,\theta}(\sfy\mid\bfx)}{q_0(\sfy\mid\bfx)}
\nonumber\\
=\;&
-\ln(1+t_\theta(\bfx))
+
\frac{\sigma_{\alpha_{\rms}}^2}{\sigma_{\rms}^4}
\frac{|\bfg_\theta^\dagger(\bfx)\sfy|^2}{1+t_\theta(\bfx)}.
\label{eq:LLR_theta_occ_rewrite}
\end{align}
After marginalizing the nuisance angle, the optimal test is
\begin{equation}
\ln\frac{P_A(1)}{P_A(0)}
+
\ln\left[
\int\rme^{\ell(\theta;\sfy\mid\bfx)}\,\rmd P_\Theta(\theta)
\right]
\mathop{\gtrless}_{\hat A=0}^{\hat A=1}0.
\label{eq:Bayes_test_occ_rewrite}
\end{equation}

\subsection{Numerical Results}
\label{sec:numerical_occupancy}

The system parameters are summarized in Table~\ref{tab:parameters_occupancy}.

\begin{table}[t]
\caption{Parameters for the occupancy-detection numerical results.}
\label{tab:parameters_occupancy}
\centering
\small
\begin{tabular}{|l|c|}
\hline
\textbf{Parameter} & \textbf{Value}\\
\hline
A priori occupancy probability $P_A(1)$ & $1/2$\\
\hline
Number of Tx antennas $M$ & $4$\\
\hline
Number of sensing Rx antennas $N_{\rms}$ & $4$\\
\hline
Number of communication Rx antennas $N_{\rmc}$ & $4$\\
\hline
Coherence period $T$ & $16$\\
\hline
Sensing SNR $P_0/\sigma_{\rms}^2$ & $\{30,0\}$ dB\\
\hline
Communication SNR $P_0/\sigma_{\rmc}^2$ & $24$ dB\\
\hline
Scattering variance $\sigma_{\alpha_{\rms}}^2$ & $1$\\
\hline
Communication Tx steering angle $\varphi$ & $-37^\circ$\\
\hline
Target angle prior $P_\Theta$ & $\operatorname{Unif}[-\pi/2,\pi/2]$\\
\hline
\end{tabular}
\end{table}

The angular expectations are evaluated numerically, and the detector error probabilities are estimated by paired Monte Carlo simulation with a target relative standard error of $0.8\%$.

\begin{figure}[t]
\centering
\input{Figures/occupancy_singlecol_fixed_ratio.pgf}
\caption{Covariance-based RDB and TV converse bounds and achievable sensing--communication regions for binary occupancy detection at sensing SNRs of $30$ and $0$~dB and communication SNR of $24$~dB. Solid lines correspond to sensing SNR $30$~dB and dashed lines to $0$~dB. System parameters are given in Table~\ref{tab:parameters_occupancy}.}
\label{fig:RD_occupancy_detection}
\end{figure}

The results in Fig.~\ref{fig:RD_occupancy_detection} separate the roles of the two links. Changing only the sensing SNR shifts the attainable detection reliability but leaves the communication-rate endpoint unchanged. Thus, sensing noise controls how distinguishable the occupied and empty likelihoods are, whereas the communication endpoint is determined by the communication channel and its power allocation.

The TV converse uses the binary decision structure directly by converting likelihood overlap into a Bayes-error bound. The RDB instead maps the sensing experiment to mutual information and then invokes the Bernoulli--Hamming distortion--rate function. Their comparison clarifies how task-specific testing information and the general rate--distortion construction lead to different converse strengths, while confirming that the RDB applies directly to a discrete decision problem.

The achievable curves illustrate the deterministic--random signaling tradeoff. Gaussian signaling is naturally favored near the communication-optimal endpoint because it maximizes output entropy for a prescribed covariance. The structured semi-unitary benchmark regularizes the transmitted energy across the sensing subspace and reduces waveform-to-waveform illumination fluctuations, thereby improving detection performance at a modest communication cost. This behavior motivates time sharing or intermediate structured distributions when neither endpoint alone provides the desired operating point.

\section{Nakagami Target-Response Estimation}
\label{sec:Nakagami}
We next consider target-response matrix estimation under non-Gaussian Nakagami fading. The Nakagami distribution is widely used to model amplitude fluctuations in multipath channels and controls fading severity through its shape parameter $m$~\cite{alouini1997capacity}. In particular, $m=1$ corresponds to Rayleigh fading, whereas $m<1$ represents more severe fading.

Channel-matrix estimation is a standard sensing abstraction in ISAC~\cite{ISAC_Caire}. If a physical parameter vector $\boldsymbol{\eta}$ generates the sensing response and the observation depends on $\boldsymbol{\eta}$ only through $\bfH_{\rms}$, then, conditioned on the known waveform,
\begin{equation}
\boldsymbol{\eta}\longrightarrow\bfH_{\rms}\longrightarrow\bfY_{\rms}
\qquad\text{given }\bfX.
\label{eq:physical_parameter_channel_markov}
\end{equation}
Thus, the target-response matrix is the information-bearing intermediate representation delivered to the sensing receiver. In this section we take recovery of $\bfH_{\rms}$ itself as the sensing task. Translating a channel-matrix MSE benchmark into an error bound for a particular physical parameter requires the corresponding forward model and is outside the scope of this example.

This continuous non-Gaussian example complements the binary occupancy problem of Section~\ref{sec:occupancy_detection}. It permits a direct comparison between the RDB and Bayesian Fisher-information methods and includes a severe-fading regime in which the conventional BCRB yields only the trivial zero lower bound while the entropy-based RDB remains finite.

\subsection{ISAC Model}
\label{sec:ISAC_model_Nakagami}
The sensing parameter is the target-response matrix itself,
\begin{equation}
\bfA=\bfH_{\rms}.
\end{equation}
The sensing and communication channels are independent,
\begin{equation}
\bfH_{\rms}\independent\bfH_{\rmc},
\label{eq:Hc_Hs_independent_Nakagami}
\end{equation}
and the observations are
\begin{equation}
\bfY_{\star}=\bfH_{\star}\bfX+\bfZ_{\star},
\qquad
\star\in\{\rms,\rmc\},
\label{eq:gaussian_channels}
\end{equation}
where the entries of $\bfZ_{\star}\in\bbC^{N_{\star}\times T}$ are i.i.d. $\cC\cN(0,\sigma_{\star}^{2})$.

The entries of each channel matrix are i.i.d. according to
\begin{equation}
H_{\star,ij}
\sim
{\cC\text{Nakagami}}(m_{\star},\omega_{\star}),
\qquad
\star\in\{\rms,\rmc\}.
\label{eq:complex_Nakagami_channels}
\end{equation}
Here, $H\sim{\cC\text{Nakagami}}(m,\omega)$ denotes a circularly symmetric complex Nakagami random variable
\begin{equation}
H=R\rme^{j\Theta},
\end{equation}
where
\begin{equation}
R\sim\operatorname{Nakagami}(m,\omega),
\qquad
\Theta\sim\operatorname{Unif}[0,2\pi),
\end{equation}
independently, and $\bbE[|H|^2]=\omega$. Its density over $\bbC$ is
\begin{equation}
f_H(h)
=
\frac{m^m}{\pi\Gamma(m)\omega^m}
|h|^{2m-2}
\exp\!\left(-\frac{m}{\omega}|h|^2\right).
\label{eq:complex_Nakagami_density}
\end{equation}

In this case study, perfect communication-channel state information is available at both the transmitter and the communication receiver. The transmitter can therefore adapt its signaling law according to $P_{\bfX\mid\bfH_{\rmc}}$. Since $\bfH_{\rms}\independent\bfH_{\rmc}$ and the transmitted waveform depends on the communication channel but not on the sensing channel,
\begin{equation}
\bfH_{\rms}\independent\bfX.
\label{eq:Hs_X_independent_Nakagami}
\end{equation}

The transmit block is subject to the short-term conditional block-power constraint
\begin{equation}
\operatorname{tr}\left(
\bbE[\bfX\bfX^\dagger\mid\bfH_{\rmc}=\bfh_{\rmc}]
\right)
\le MTP_0,
\qquad\forall\,\bfh_{\rmc}.
\label{eq:APowerConstraintShortTerm}
\end{equation}
Throughout this section, $\cC_{\rm N}$ denotes the analogue of the capacity--distortion region in Definition~\ref{def:capacity} when the admissible signaling objects are conditional input laws $P_{\bfX\mid\bfH_{\rmc}}$ satisfying~\eqref{eq:APowerConstraintShortTerm}.

The sensing distortion is
\begin{equation}
D
=
\bbE\left[
\|\bfH_{\rms}-\widehat{\bfH}_{\rms}\|_F^2
\right].
\label{eq:Nakagami_MSE_definition}
\end{equation}
The minimum distortion is $\operatorname{MMSE}(\bfH_{\rms}\mid\bfY_{\rms},\bfX)$, attained by the posterior-mean estimator. A scalar derivation for general circularly symmetric fading is provided in Appendix~\ref{app:nakagami_scalar_mmse}; its complexity motivates the information-theoretic converse used for the MIMO model.

\subsection{Bayesian Fisher Information and Nakagami RDB}
\label{subsec:Nakagami_intermediate}

\begin{theorem}[Fisher information of the complex Nakagami prior]
\label{thm:Fisher_info_Nakagami}
Let $\sfH\sim{\cC\text{Nakagami}}(m,\omega)$ with $m\ge1/2$, and define
\begin{equation}
J_{\sfH}
\triangleq
\bbE\left[
\left|\frac{\partial}{\partial\sfH^*}\ln f_{\sfH}(\sfH)\right|^2
\right].
\end{equation}
Then
\begin{equation}
J_{\sfH}
=
\begin{cases}
\dfrac{m}{\omega},&m\ge1,\\[2mm]
+\infty,&\dfrac12\le m<1.
\end{cases}
\label{eq:FIM_prior_Nakagami}
\end{equation}
\end{theorem}

\begin{proof}
From~\eqref{eq:complex_Nakagami_density},
\begin{equation}
\frac{\partial}{\partial h^*}\ln f_H(h)
=
h\left(\frac{m-1}{|h|^2}-\frac{m}{\omega}\right).
\label{eq:Nakagami_score}
\end{equation}
Thus, with $R=|H|$,
\begin{equation}
\left|\frac{\partial}{\partial H^*}\ln f_H(H)\right|^2
=
\frac{(m-1)^2}{R^2}
-
\frac{2m(m-1)}{\omega}
+
\frac{m^2}{\omega^2}R^2.
\label{eq:Nakagami_score_squared}
\end{equation}
For $R\sim\operatorname{Nakagami}(m,\omega)$, $\bbE[R^2]=\omega$, while $\bbE[R^{-2}]=m/[\omega(m-1)]$ is finite only for $m>1$. Substitution gives $J_{\sfH}=m/\omega$ for $m>1$. At $m=1$, the singular term vanishes and the same expression holds. For $m<1$, $\bbE[R^{-2}]$ diverges, yielding $J_{\sfH}=+\infty$.
\end{proof}

In Bayesian estimation, the information matrix combines the contribution of the observation likelihood with the Fisher information of the parameter distribution. Theorem~\ref{thm:Fisher_info_Nakagami} evaluates the latter contribution. For $m<1$, it diverges because the Nakagami density is singular at the origin, and the regularity conditions underlying the standard BCRB are not satisfied; the resulting lower bound reduces to zero. The Nakagami differential entropy remains finite, so the RDB continues to provide a nontrivial converse.

Define
\begin{equation}
\kappa_m \triangleq \frac{\Gamma(m)}{m} \exp\!\left((m-1)(1-\psi(m))\right).
\label{eq:kappa_m_Nakagami}
\end{equation}

\begin{prop}[Entropy-power RDB for Nakagami fading]
\label{prop:inverse_RDfunc_Nakagami}
Let $\bfV\in\bbC^n$ have i.i.d. ${\cC\text{Nakagami}}(m,\omega)$ entries and squared-error distortion. Then, for $r$ measured in bits,
\begin{equation}
D_{\bfV,d}(r)
\ge
n\omega\kappa_m \, 2^{-r/n},
\qquad r\ge0.
\label{eq:inverse_RD_Nakagamivector_complex}
\end{equation}
\end{prop}

\begin{proof}
For one complex Nakagami random variable,
\begin{equation}
h(H)
=
\frac{1}{\ln2}
\left[
\ln\left(\frac{\pi\Gamma(m)\omega}{m}\right)
+m+(1-m)\psi(m)
\right]
\label{eq:Nakagami_entropy_bits}
\end{equation}
bits. Since the entries of $\bfV$ are independent, $h(\bfV)=nh(H)$. Applying~\eqref{eq:SLB_inv_complex} yields~\eqref{eq:inverse_RD_Nakagamivector_complex}. 
% \begin{equation}
% D_{\bfV,d}(r)
% \ge
% n\omega\rme^{c_m} \, 2^{-r/n}.
% \end{equation}
\end{proof}
Proposition~\ref{prop:inverse_RDfunc_Nakagami} is an SLB specialization of the distortion--rate function; it does not require the exact Nakagami rate--distortion function.

\subsection{RDB- and BCRB-Based Sensing Bounds}
\label{subsec:Nakagami_sensing_bounds}

\begin{prop}[RDB-based MMSE lower bound]
\label{thm:MMSE_bound_Nakagami}
For any $m_{\rms}\ge1/2$,
\begin{align}
\operatorname{MMSE}(\bfH_{\rms}&\mid\bfY_{\rms},\bfX)
\ge 
MN_{\rms}\omega_{\rms}\kappa_{m_{\rms}}
\nonumber\\
&\times
\bbE\left[
\det\left(
\bfI_M+
\frac{\omega_{\rms}}{\sigma_{\rms}^{2}}
\bfX\bfX^\dagger
\right)^{-1/M}
\right].
\label{eq:lower_bound_MMSE_Nakagami}
\end{align}
Moreover, under~\eqref{eq:APowerConstraintShortTerm},
\begin{equation}
\operatorname{MMSE}(\bfH_{\rms}\mid\bfY_{\rms},\bfX)
\ge
\frac{MN_{\rms}\omega_{\rms}\kappa_{m_{\rms}}}
{1+\frac{\omega_{\rms}}{\sigma_{\rms}^{2}}TP_0}.
\label{eq:best_lower_bound_MMSE_Nakagami}
\end{equation}
\end{prop}

\begin{proof}
For fixed $\bfX=\bfx$, Proposition~\ref{prop:inverse_RDfunc_Nakagami} and Corollary~\ref{cor:waveform_conditioned_RDB} give
{\small
\begin{equation}
\bbE[\|\bfH_{\rms}-\widehat{\bfH}_{\rms}\|_F^2\mid\bfX=\bfx]
\ge
MN_{\rms}\omega_{\rms}\kappa_{m_{\rms}}
2^{-I(\bfH_{\rms};\bfY_{\rms}\mid\bfX=\bfx)/(MN_{\rms})}.
\label{eq:Nakagami_RDB_before_MI_bound}
\end{equation}
}
Gaussian maximization of differential entropy for fixed covariance gives
\begin{align}
I(\bfH_{\rms};\bfY_{\rms}\mid\bfX=\bfx)
&\le
N_{\rms}
\log\det\left(
\bfI_T+
\frac{\omega_{\rms}}{\sigma_{\rms}^{2}}
\bfx^\dagger\bfx
\right)
\nonumber\\
&=
N_{\rms}
\log\det\left(
\bfI_M+
\frac{\omega_{\rms}}{\sigma_{\rms}^{2}}
\bfx\bfx^\dagger
\right),
\label{eq:upper_I_HY_given_x}
\end{align}
where the last equality follows from Sylvester's determinant identity. Substitution and averaging over $\bfX$ give~\eqref{eq:lower_bound_MMSE_Nakagami}. The determinant--trace inequality
\begin{equation}
\det(\bfB)^{-1/M}\ge\frac{M}{\operatorname{tr}(\bfB)},\qquad\bfB\succ{\bf0},
\end{equation}
followed by Jensen's inequality and~\eqref{eq:APowerConstraintShortTerm}, yields~\eqref{eq:best_lower_bound_MMSE_Nakagami}.
\end{proof}

\begin{prop}[BCRB-based MMSE lower bound]
\label{thm:MMSE_BCRB_Nakagami}
For $m_{\rms}\ge1$,
\begin{align}
\operatorname{MMSE}(\bfH_{\rms}\mid\bfY_{\rms},\bfX)
&\ge
N_{\rms}\frac{\omega_{\rms}}{m_{\rms}}
%\nonumber\\&\times
\bbE\left[
\operatorname{tr}\left(
\bfI_M+
\frac{\omega_{\rms}}{m_{\rms}\sigma_{\rms}^{2}}
\bfX\bfX^\dagger
\right)^{-1}
\right].
\label{eq:BCRB_Nakagami}
\end{align}
Under~\eqref{eq:APowerConstraintShortTerm},
\begin{equation}
\operatorname{MMSE}(\bfH_{\rms}\mid\bfY_{\rms},\bfX)
\ge
\frac{MN_{\rms}\omega_{\rms}}
{m_{\rms}+\frac{\omega_{\rms}}{\sigma_{\rms}^{2}}TP_0}.
\label{eq:best_lower_bound_BCRB_MMSE_Nakagami}
\end{equation}
\end{prop}

\begin{proof}
Conditioned on $\bfX$, the Bayesian information matrix for each sensing-channel row is
\begin{equation}
\frac{m_{\rms}}{\omega_{\rms}}\bfI_M
+
\frac{1}{\sigma_{\rms}^2}\bfX\bfX^\dagger.
\end{equation}
The Bayesian Cram\'er--Rao inequality and summation over the $N_{\rms}$ independent rows yield~\eqref{eq:BCRB_Nakagami}. Let
\begin{align}
\bfB=(m_{\rms}/\omega_{\rms})\bfI_M+(1/\sigma_{\rms}^2)\bfX\bfX^\dagger. 
\end{align}
Applying the arithmetic--geometric mean inequality separately to the eigenvalues of $\bfB$ and their reciprocals gives $\operatorname{tr}(\bfB^{-1})\ge M^2/\operatorname{tr}(\bfB)$. Jensen's inequality and~\eqref{eq:APowerConstraintShortTerm} then yield~\eqref{eq:best_lower_bound_BCRB_MMSE_Nakagami}.
\end{proof}

\begin{rem}[RDB--BCRB comparison at the sensing endpoint]
\label{rem:Nakagami_RDB_BCRB}
Let
\begin{equation}
u\triangleq\frac{\omega_{\rms}P_0T}{\sigma_{\rms}^{2}}.
\end{equation}
Then
\begin{equation}
\epsilon_{\rm RDB}
=
\frac{MN_{\rms}\omega_{\rms}\kappa_{m_{\rms}}}{1+u},
\qquad
\epsilon_{\rm BCRB}
=
\frac{MN_{\rms}\omega_{\rms}}{m_{\rms}+u}.
\label{eq:Nakagami_endpoint_bounds}
\end{equation}
For $m_{\rms}=1$, $\kappa_1=1$ and the two expressions coincide. For $m_{\rms}>1$, $\epsilon_{\rm RDB}\ge\epsilon_{\rm BCRB}$ if and only if
\begin{equation}
u
\le
u^\star(m_{\rms})
\triangleq
\frac{m_{\rms}\kappa_{m_{\rms}}-1}{1-\kappa_{m_{\rms}}}.
\label{eq:Nakagami_SNR_threshold}
\end{equation}
Thus, when both bounds are available, their ordering depends on the fading parameter and sensing SNR. For $m_{\rms}<1$, the Fisher-information contribution of the parameter distribution is not finite and the conventional Fisher-information expression degenerates to zero, while the regularity conditions required for the standard BCRB are not satisfied, whereas the RDB remains finite.
\end{rem}

\subsection{Covariance-Based ISAC Converse}
\label{subsec:Nakagami_covariance_converse}
Define
\begin{equation}
\bfQ(\bfh_{\rmc})
\triangleq
\bbE[\bfX\bfX^\dagger\mid\bfH_{\rmc}=\bfh_{\rmc}]
\label{eq:Q_Hc_Nakagami}
\end{equation}
and
\begin{equation}
\cQ_{\rm N}
\triangleq
\left\{
\bfQ(\cdot):
\bfQ(\bfh_{\rmc})\succeq{\bf0},\;
\operatorname{tr}\bfQ(\bfh_{\rmc})\le MTP_0,\;
\forall\bfh_{\rmc}
\right\}.
\label{eq:Qset_Nakagami}
\end{equation}

\begin{prop}[Communication-rate covariance bound]
\label{prop:upper_ergodic_info_rate}
For any admissible $P_{\bfX\mid\bfH_{\rmc}}$, by the Gaussian maximum-entropy argument underlying the MIMO capacity expression~\cite{telatar1999capacity},
{%\small
\begin{align}
\frac1T I(\bfX;\bfY_{\rmc}\mid\bfH_{\rmc})
\le 
\bbE_{\bfH_{\rmc}} \hspace*{-.2cm}
\left[
\log\det\left(
\bfI_{N_{\rmc}}
+
\frac{1}{T\sigma_{\rmc}^{2}}
\bfH_{\rmc}\bfQ(\bfH_{\rmc})\bfH_{\rmc}^\dagger
\right)
\right].
\label{eq:upper_bound_ergodic_comm_rate}
\end{align}
}
Equality is attained by conditionally Gaussian signaling with independent columns and per-column covariance $\bfQ(\bfH_{\rmc})/T$.
\end{prop}

\begin{proof}
Condition on $\bfH_{\rmc}=\bfh_{\rmc}$. Subadditivity of output entropy and Gaussian maximum entropy give
\begin{equation}
I(\bfX;\bfY_{\rmc}\mid\bfh_{\rmc})
\le
\sum_{t=1}^{T}
\log\det\!\left(
\bfI_{N_{\rmc}}+
\frac{1}{\sigma_{\rmc}^2}
\bfh_{\rmc}\bfQ_t\bfh_{\rmc}^\dagger
\right),
\end{equation}
where $\bfQ_t=\bbE[\bfX_t\bfX_t^\dagger\mid\bfh_{\rmc}]$ and $\sum_t\bfQ_t=\bfQ(\bfh_{\rmc})$. Concavity of $\log\det(\cdot)$ bounds the sum by $T$ times its value at the average covariance $\bfQ(\bfh_{\rmc})/T$. Averaging over $\bfH_{\rmc}$ proves~\eqref{eq:upper_bound_ergodic_comm_rate}; independent Gaussian columns attain every step.
\end{proof}

Define
\begin{align}
\underline D_{\rm N}(\bfQ)
\triangleq\;&
MN_{\rms}\omega_{\rms}\kappa_{m_{\rms}}
\bbE_{\bfH_{\rmc}}
\left[
\det\left(
\bfI_M+
\frac{\omega_{\rms}}{\sigma_{\rms}^{2}}
\bfQ(\bfH_{\rmc})
\right)^{-1/M}
\right],
\label{eq:Dbar_Nakagami}\\
\overline R_{\rm N}(\bfQ)
\triangleq\;&
\bbE_{\bfH_{\rmc}}
\left[
\log\det\left(
\bfI_{N_{\rmc}}+
\frac{1}{T\sigma_{\rmc}^{2}}
\bfH_{\rmc}\bfQ(\bfH_{\rmc})\bfH_{\rmc}^\dagger
\right)
\right].
\label{eq:Rbar_Nakagami}
\end{align}

\begin{theorem}[Covariance-based Nakagami converse]
\label{thm:converse_Nakagami}
For the model of Section~\ref{sec:ISAC_model_Nakagami},
\begin{equation}
\cC_{\rm N}
\subseteq
\overline{\operatorname{conv}}
\bigcup_{\bfQ(\cdot)\in\cQ_{\rm N}}
\left\{
(R,D):R\le\overline R_{\rm N}(\bfQ),\;D\ge\underline D_{\rm N}(\bfQ)
\right\}.
\label{eq:enlarged_outer}
\end{equation}
\end{theorem}

\begin{proof}
Since $\bfB\mapsto\det(\bfB)^{-1/M}$ is convex over the positive-definite cone, conditional Jensen's inequality applied to~\eqref{eq:lower_bound_MMSE_Nakagami} gives $D\ge\underline D_{\rm N}(\bfQ)$. Proposition~\ref{prop:upper_ergodic_info_rate} gives $R\le\overline R_{\rm N}(\bfQ)$. Taking the union over feasible covariance mappings and its convex closure completes the proof.
\end{proof}

For numerical evaluation, the Pareto boundary is traced, for each $\bfh_{\rmc}$ and $\lambda\ge0$, by
{\footnotesize
\begin{align}
\bfQ_\lambda(\bfh_{\rmc})
\in
\argmin_{\bfQ\succeq{\bf0}:\,\operatorname{tr}(\bfQ)\le MTP_0}
\Bigg\{&
MN_{\rms}\omega_{\rms}\kappa_{m_{\rms}}
\det\left(
\bfI_M+
\frac{\omega_{\rms}}{\sigma_{\rms}^{2}}\bfQ
\right)^{-1/M}
\nonumber\\
&-
\lambda
\log\det\left(
\bfI_{N_{\rmc}}+
\frac{1}{T\sigma_{\rmc}^{2}}
\bfh_{\rmc}\bfQ\bfh_{\rmc}^\dagger
\right)
\Bigg\}.
\label{eq:optimization_Qlambda}
\end{align}
}

\subsection{Achievable Signaling Strategies}
\label{subsec:Nakagami_achievable}
We compare the converse with Gaussian and semi-unitary signaling. Both use the LMMSE estimator, whose MSE is
\begin{equation}
D_{\rm L}
=
N_{\rms}\omega_{\rms}
\bbE\left[
\operatorname{tr}\left(
\bfI_M+
\frac{\omega_{\rms}}{\sigma_{\rms}^{2}}
\bfX\bfX^\dagger
\right)^{-1}
\right].
\label{eq:lmmse_mse_general}
\end{equation}
For $m_{\rms}=1$, the sensing channel is circularly Gaussian and LMMSE coincides with MMSE. For $m_{\rms}\neq1$,~\eqref{eq:lmmse_mse_general} remains achievable but is generally not MMSE-optimal.

\subsubsection{Gaussian Signaling}
Conditioned on $\bfH_{\rmc}=\bfh_{\rmc}$,
\begin{equation}
\bfX_t
\sim
\cC\cN({\bf0},\boldsymbol{\Sigma}_{\rm G}(\bfh_{\rmc})),
\qquad t\in[T],
\label{eq:Nakagami_Gaussian_signaling}
\end{equation}
independently, with $\operatorname{tr}(\boldsymbol{\Sigma}_{\rm G})\le MP_0$. The communication rate is
\begin{equation}
R_{\rm G}
=
\bbE_{\bfH_{\rmc}}
\left[
\log\det\left(
\bfI_{N_{\rmc}}+
\frac{1}{\sigma_{\rmc}^{2}}
\bfH_{\rmc}\boldsymbol{\Sigma}_{\rm G}(\bfH_{\rmc})\bfH_{\rmc}^\dagger
\right)
\right].
\label{eq:Nakagami_Gaussian_rate}
\end{equation}
A family of operating points is generated by
{\footnotesize
\begin{align}
\boldsymbol{\Sigma}_{{\rm G},\lambda}^{\star}(\bfh_{\rmc})
\in
\argmin_{\boldsymbol{\Sigma}\succeq{\bf0}:\,\operatorname{tr}(\boldsymbol{\Sigma})\le MP_0}
\Bigg\{&
N_{\rms}\omega_{\rms}
\operatorname{tr}\left(
\bfI_M+
\frac{\omega_{\rms}}{\sigma_{\rms}^{2}}
T\boldsymbol{\Sigma}
\right)^{-1}
\nonumber\\
&-
\lambda
\log\det\left(
\bfI_{N_{\rmc}}+
\frac{1}{\sigma_{\rmc}^{2}}
\bfh_{\rmc}\boldsymbol{\Sigma}\bfh_{\rmc}^\dagger
\right)
\Bigg\}.
\label{eq:optimization_SigmaG}
\end{align}
}
The first term is a covariance-based design surrogate. After the covariance is selected, the actual achievable sensing distortion is evaluated from~\eqref{eq:lmmse_mse_general} using
\begin{equation}
\bfX
=
\left(\boldsymbol{\Sigma}_{{\rm G},\lambda}^{\star}(\bfH_{\rmc})\right)^{1/2}\bfG,
\qquad
G_{ij}\sim\cC\cN(0,1).
\label{eq:Nakagami_Gaussian_MC}
\end{equation}

\subsubsection{Semi-Unitary Signaling}
Assume $T\ge M$, as in the numerical configurations considered below, and let
\begin{equation}
\bfX=\sqrt{TP_0}\,\bfS,
\label{eq:semi_unitary_signalling}
\end{equation}
where $\bfS$ is uniformly distributed on the complex Stiefel manifold
\begin{equation}
\cV_{M,T}
=
\{\bfS\in\bbC^{M\times T}:\bfS\bfS^\dagger=\bfI_M\}.
\end{equation}
Thus,
\begin{equation}
\bfX\bfX^\dagger=TP_0\bfI_M
\label{eq:semiunitary_gram}
\end{equation}
for every waveform realization, and
\begin{equation}
D_{\rm U}
=
\frac{MN_{\rms}\omega_{\rms}}
{1+\frac{\omega_{\rms}}{\sigma_{\rms}^{2}}TP_0}.
\label{eq:lmmse_mse_semiunitary}
\end{equation}

For communication, we use the high-SNR approximation in~\cite{ISAC_Caire}. Define
\begin{equation}
M_{\rm U}
\triangleq
\operatorname{rank}(\bfH_{\rmc}\bfH_{\rmc}^\dagger).
\label{eq:MU_Nakagami}
\end{equation}
Since the entries of $\bfH_{\rmc}$ have continuous distributions,
\begin{equation}
M_{\rm U}=\min(M,N_{\rmc})
\quad\text{a.s.}
\label{eq:MU_rank_Nakagami}
\end{equation}
For the numerical configurations below, $N_{\rmc}=2<M$, hence
\begin{align}
M_{\rm U}=2.
\label{eq:MU_numeric_Nakagami}
\end{align}
In this full-row-rank case, the corresponding rate approximation is
\begin{align}
R_{\rm U}
\approx\;&
\left(1-\frac{M_{\rm U}}{2T}\right)
\bbE_{\bfH_{\rmc}}
\left[
\log\det\left(
\frac{P_0}{\sigma_{\rmc}^{2}}
\bfH_{\rmc}\bfH_{\rmc}^\dagger
\right)
\right]
+
c_{\rm U},
\label{eq:Nakagami_semiunitary_rate}
\end{align}
where
\begin{equation}
c_{\rm U}
=
\frac{M_{\rm U}}{T}
\left[
\left(T-\frac{M_{\rm U}}{2}\right)
\log\left(\frac{T}{\rme}\right)
+
\log\left(\frac{2\sqrt\pi}{\Gamma(T)}\right)
\right].
\label{eq:cU_Nakagami}
\end{equation}
We therefore interpret the semi-unitary curve as a high-SNR achievable benchmark.

\subsubsection{Time Sharing}
Time sharing between achievable operating points $(R_1,D_1)$ and $(R_2,D_2)$ produces
\begin{align}
R_\tau&=\tau R_1+(1-\tau)R_2,
\nonumber\\
D_\tau&=\tau D_1+(1-\tau)D_2,
\qquad0\le\tau\le1.
\label{eq:Nakagami_time_sharing}
\end{align}
The interpolation is performed in the original linear $(D,R)$ coordinates; the resulting points are only afterward displayed on the logarithmic distortion axis.

\subsection{Numerical Results}
\label{sec:numerical_Nakagami}
We consider
\begin{equation}
m_{\rms}=m_{\rmc}\in\{0.5,1,2\},
\end{equation}
with $\omega_{\rms}=1$ and $\omega_{\rmc}=M^{-1}$. The converse and achievable regions are evaluated from the covariance optimizations above; time sharing is performed in linear $(D,R)$ coordinates before the logarithmic distortion axis is applied.

\begin{table}[t]
\caption{Parameters for the Nakagami target-response estimation results.}
\label{tab:params_Nakagami}
\centering
\small
\resizebox{\columnwidth}{!}{%
\begin{tabular}{|l|c|c|}
\hline
\textbf{Parameter}&\textbf{Fig.~\ref{fig:RD_m_all_M_16}}&\textbf{Fig.~\ref{fig:RD_m_all_M_8}}\\
\hline
Number of Tx antennas $M$&$16$&$8$\\
\hline
Number of sensing Rx antennas $N_{\rms}$&$4$&$4$\\
\hline
Number of communication Rx antennas $N_{\rmc}$&$2$&$2$\\
\hline
Coherence period $T$&$4M=64$&$2M=16$\\
\hline
Sensing SNR $P_0/\sigma_{\rms}^{2}$&$24$ dB&$5$ dB\\
\hline
Communication SNR $P_0/\sigma_{\rmc}^{2}$&$24$ dB&$15$ dB\\
\hline
Sensing scale $\omega_{\rms}$&$1$&$1$\\
\hline
Communication scale $\omega_{\rmc}$&$M^{-1}$&$M^{-1}$\\
\hline
Nakagami parameters&$m_{\rms}=m_{\rmc}\in\{0.5,1,2\}$&$m_{\rms}=m_{\rmc}\in\{0.5,1,2\}$\\
\hline
Semi-unitary rank $M_{\rm U}$&$2$&$2$\\
\hline
\end{tabular}%
}
\end{table}

\begin{figure}[t]
\centering
\input{Figures/RDB_M16_twocol.pgf}
\caption{RDB converse and achievable sensing--communication regions for Nakagami target-response estimation with $M=16$, $N_{\rms}=4$, $N_{\rmc}=2$, $T=64$, and sensing and communication SNRs of $24$~dB. The RDB converse uses the SLB on the Nakagami distortion--rate function. Solid lines denote the RDB converse, dash-dotted lines Gaussian signaling, dashed lines the semi-unitary benchmark, and dotted lines time sharing between Gaussian and semi-unitary signaling.}
\label{fig:RD_m_all_M_16}
\end{figure}

\begin{figure}[t]
\centering
\input{Figures/RDB_M8_twocol.pgf}
\caption{RDB converse and achievable sensing--communication regions for Nakagami target-response estimation with $M=8$, $N_{\rms}=4$, $N_{\rmc}=2$, $T=16$, sensing SNR $5$~dB, and communication SNR $15$~dB. The RDB converse uses the SLB on the Nakagami distortion--rate function. The line styles are the same as in Fig.~\ref{fig:RD_m_all_M_16}.}
\label{fig:RD_m_all_M_8}
\end{figure}

Figs.~\ref{fig:RD_m_all_M_16} and~\ref{fig:RD_m_all_M_8} show that Gaussian signaling nearly reaches the communication endpoint of the covariance outer region in both operating regimes. This identifies the communication-side Gaussian covariance relaxation as a minor source of the remaining gap. Away from that endpoint, the separation is governed primarily by the sensing relaxations and by the estimator used in the achievable construction.

Semi-unitary signaling fixes the sample Gram matrix, eliminates block-to-block covariance fluctuations, and illuminates the transmit dimensions uniformly. These properties improve LMMSE sensing performance while retaining a substantial communication rate. Gaussian and semi-unitary signaling therefore represent the two ends of the deterministic--random tradeoff: the former preserves communication-bearing randomness, while the latter imposes structure that is favorable to sensing. Time sharing supplies useful intermediate operating points.

The Rayleigh case $m=1$ provides a structural consistency check. The sensing prior is then circular Gaussian, the complex entropy-power expression is exact for the relevant covariance, and LMMSE coincides with MMSE. Accordingly, the semi-unitary sensing endpoint meets the RDB endpoint. For $m\ne1$, the residual gap can contain several effects: the Shannon lower-bound relaxation of the Nakagami distortion--rate function, the Gaussian maximum-entropy upper bound on sensing mutual information, and the suboptimality of LMMSE for a non-Gaussian prior. %In the severe-fading case $m=0.5$, the standard BCRB yields only the trivial zero lower bound, whereas the entropy-based RDB remains finite and informative.

\section{Conclusion}
\label{sec:conclusion}
We developed a rate--distortion converse for the sensing--communication tradeoff in ISAC. The general conditional RDB yields a waveform-conditioned ISAC bound that applies the distortion--rate function before averaging over transmitted waveform realizations. This preserves waveform-dependent sensing information and produces a converse that is no weaker than its averaged-information counterpart. The framework accommodates general losses and random conditional sensing laws without imposing regularity conditions on the parameter distribution. It is exact in the high-sensing-noise limit, and its scalar entropy-power specialization is no looser than the BCRB.

Binary occupancy detection and Nakagami target-response estimation demonstrated the scope of the framework. The occupancy example established direct applicability to discrete sensing under Hamming loss and nuisance-induced mixture observations, while the Nakagami example remained informative in a severe-fading regime where the standard BCRB yields only the trivial zero lower bound. Across both examples, Gaussian signaling was favored near the communication endpoint, whereas semi-unitary structure improved sensing by controlling waveform covariance fluctuations. Future work will focus on developing achievability schemes and extending the RDB framework to multi-user and multi-target ISAC models.

\appendices

\section{General Shannon Lower Bounds}
\label{app:general_SLB}
The main text uses the squared-error Shannon lower bounds required by the Nakagami example, while the occupancy example uses the exact Bernoulli--Hamming distortion--rate function. For completeness, we record here broader continuous and discrete Shannon lower bounds that motivate the generic relaxation in~\eqref{eq:C_R_RSLB}. All entropies and rates in this appendix are measured in bits.

\subsection{Continuous Sources with Norm-Power Distortion}
\label{app:general_continuous_SLB}
Let $\sfV\in\bbR^n$ have a continuous distribution with finite differential entropy and consider
\begin{equation}
d(\sfv,\hat{\sfv})=\|\sfv-\hat{\sfv}\|_{\ell}^{k},\qquad k>0,\quad 1\le \ell\le\infty,
\label{eq:general_norm_power_distortion}
\end{equation}
where $V_{n,\ell}$ denotes the volume of the unit ball of $\|\cdot\|_{\ell}$ in $\bbR^n$. The classical Shannon lower bound gives
\begin{align}
R_{\sfV,d}(D)
&\ge
R_{\sfV,{\rm SLB}}(D)
\nonumber\\
&\triangleq
h(\sfV)
+\log\!\left(\frac{1}{V_{n,\ell}\Gamma(n/k+1)}\right)
\nonumber\\
&\quad
+\frac{n}{k}\log\!\left(\frac{n}{kD\rme}\right).
\label{eq:general_continuous_SLB_R}
\end{align}
and hence
\begin{equation}
D_{\sfV,d}(r)
\ge
\underline D_{\sfV,d}^{\rm SLB}(r)
\triangleq
\frac{n}{k\rme\left[V_{n,\ell}\Gamma(n/k+1)\right]^{k/n}}
2^{\frac{k}{n}(h(\sfV)-r)}.
\label{eq:general_continuous_SLB_D}
\end{equation}
For $k=\ell=2$, $V_{n,2}=\pi^{n/2}/\Gamma(n/2+1)$ and~\eqref{eq:general_continuous_SLB_D} reduces to the real squared-error expression in~\eqref{eq:SLB_inv_continuous}; the corresponding proper-complex specialization is~\eqref{eq:SLB_inv_complex}.

\subsection{Discrete Sources with Permutation Distortion Matrices}
\label{app:general_discrete_SLB}
Let $\sfV$ be a discrete source and assume that every column of its distortion matrix is a permutation of the same multiset $\{d_1,\ldots,d_m\}$. Define
\begin{equation}
\phi(D)\triangleq\max_{\bfp:\,\sum_{i=1}^{m}p_i d_i\le D}H(\bfp),
\label{eq:phi_discrete_SLB}
\end{equation}
which is non-decreasing and concave in $D$. The discrete Shannon lower bound~\cite[Prob.~10.6]{Cover:2006} is
\begin{equation}
R_{\sfV,d}(D)\ge H(\sfV)-\phi(D).
\label{eq:general_discrete_SLB_R}
\end{equation}
For the non-decreasing function $\phi$, define the generalized inverse
\begin{equation}
\phi^{\leftarrow}(s)\triangleq\inf\{D:\phi(D)\ge s\}.
\label{eq:phi_generalized_inverse}
\end{equation}
Then
\begin{equation}
D_{\sfV,d}(r)
\ge
\underline D_{\sfV,d}^{\rm SLB}(r)
\triangleq
\phi^{\leftarrow}\!\left(H(\sfV)-r\right),\qquad0\le r\le H(\sfV).
\label{eq:general_discrete_SLB_D}
\end{equation}
Since $\phi$ is non-decreasing, the generalized inverse in~\eqref{eq:phi_generalized_inverse} is taken over the set $\phi(D)\ge H(\sfV)-r$.

More general Shannon-type lower bounds can be formulated for random variables supported on general sets and measures. Among the most general formulations relevant here,~\cite{riegler2018rate} develops a rate--distortion lower bound for general source distributions under a broad subregularity condition on an underlying $\sigma$-finite reference measure, encompassing Euclidean sources as well as distributions supported on manifolds and fractal sets. These more general forms can be inserted into~\eqref{eq:generic_DSLB_definition} without changing the waveform-conditioned ISAC argument.

\section{Proofs of Tightness and the BCRB Comparison}
\label{app:tightness_BCRB}

\subsection{High-Sensing-Noise Tightness}
\label{app:high_noise_tightness}
Define
\begin{equation}
I_\eta(\bfx)
\triangleq
I(\bfA;\bfY_{\rms}^{(\eta)}\mid\bfX=\bfx).
\end{equation}
By Corollary~\ref{cor:waveform_conditioned_RDB},
\begin{equation}
D_\eta^\star(P_{\bfX})
\ge
\bbE_{\bfX}[D_{\bfA,d}(I_\eta(\bfX))].
\label{eq:appendix_high_noise_lower}
\end{equation}
Under~\eqref{eq:high_noise_condition}, $I_\eta(\bfx)\to0$ for $P_{\bfX}$-almost every $\bfx$. Since $D_{\bfA,d}(0^+)<\infty$ and
\begin{equation}
0\le D_{\bfA,d}(r)\le D_{\bfA,d}(0^+)=\inf_{\bfc}\bbE[d(\bfA,\bfc)],
\end{equation}
the zero-rate property and dominated convergence give
\begin{equation}
\lim_{\eta\to\infty}
\bbE_{\bfX}[D_{\bfA,d}(I_\eta(\bfX))]
=
D_{\bfA,d}(0^+).
\label{eq:appendix_high_noise_RDB_limit}
\end{equation}
Conversely, the sensing receiver can ignore its observation and use an optimal constant reconstruction, so
\begin{equation}
D_\eta^\star(P_{\bfX})
\le
D_{\bfA,d}(0^+).
\label{eq:appendix_high_noise_upper}
\end{equation}
The result follows by squeezing.

\subsection{Conditional Stam Inequality}
\label{app:conditional_Stam}
For scalar $\sfV$, define
\begin{equation}
N(\sfV\mid\sfW)
\triangleq
\frac{1}{2\pi\rme}2^{2h(\sfV\mid\sfW)}
\end{equation}
and
\begin{equation}
J(\sfV\mid\sfW)
\triangleq
\bbE_{\sfW}[J(\sfV\mid\sfW=\sfw)].
\end{equation}

\begin{lemma}[Conditional Stam inequality]
\label{lem:cond_stam}
Whenever the quantities are well defined,
\begin{equation}
N(\sfV\mid\sfW)\ge J^{-1}(\sfV\mid\sfW).
\label{eq:conditional_Stam_appendix}
\end{equation}
\end{lemma}

\begin{proof}
For each $\sfW=\sfw$, Stam's inequality~\cite{stam1959some} gives
\begin{equation}
N(\sfV\mid\sfW=\sfw)J(\sfV\mid\sfW=\sfw)\ge1.
\end{equation}
Taking natural logarithms and averaging yields
\begin{equation}
\bbE_{\sfW}[\ln N(\sfV\mid\sfW)]
\ge
-\bbE_{\sfW}[\ln J(\sfV\mid\sfW)].
\end{equation}
The left side exponentiates to $N(\sfV\mid\sfW)$ because conditional entropy is the average of $h(\sfV\mid\sfW=\sfw)$. Moreover, concavity of $\ln(\cdot)$ gives
\begin{equation}
\bbE_{\sfW}[\ln J(\sfV\mid\sfW)]
\le
\ln J(\sfV\mid\sfW).
\end{equation}
Exponentiating proves~\eqref{eq:conditional_Stam_appendix}.
\end{proof}

Combining Lemma~\ref{lem:cond_stam} with the scalar squared-error SLB proves Proposition~\ref{prop:salardominance}.

\subsection{Low-Noise Gaussian Test Channel}
\label{app:low_noise_tightness}
For~\eqref{eq:GaussChannel},
\begin{align}
D_{\rm SLB}(\sigma)
&=
\frac{n}{2\pi\rme}
2^{\frac{2}{n}h(\sfV\mid\sfW)}.
\end{align}
The identity
\begin{equation}
h(\sfV\mid\sfW)
=
h(\sigma\sfZ)-I(\sfZ;\sfV+\sigma\sfZ)
\end{equation}
gives
\begin{equation}
D_{\rm SLB}(\sigma)
=
n\sigma^2
2^{-\frac{2}{n}I(\sfZ;\sfV+\sigma\sfZ)}.
\label{eq:low_noise_SLB_exact}
\end{equation}
Moreover,
\begin{equation}
I(\sfZ;\sfV+\sigma\sfZ)
=
h(\sfV+\sigma\sfZ)-h(\sfV),
\end{equation}
which converges to zero under the entropy-continuity assumption stated in Section~\ref{sec:low_noise_tightness}. Hence $D_{\rm SLB}(\sigma)/\sigma^2\to n$. Finally,
\begin{align}
D_{\rm SLB}(\sigma)
&\le
D_{\sfV,d}(I(\sfV;\sfW))
\le
\operatorname{MMSE}(\sfV\mid\sfW)
\nonumber\\
&\le
\bbE[\|\sfV-\sfW\|_2^2]
=
n\sigma^2.
\end{align}
where the last inequality uses the feasible estimator $\hat{\sfV}=\sfW$. Dividing by $\sigma^2$ and applying the squeeze theorem proves~\eqref{eq:low_noise_joint_limit}.

\section{Derivations for the Occupancy Converse}
\label{app:occupancy_details}

\subsection{Bhattacharyya Coefficient for Proper Complex Gaussians}
\label{app:occupancy_bhattacharyya}
Let
\begin{equation}
p_i(\sfy)=\cC\cN({\bf0},\boldsymbol{\Sigma}_i),\qquad i\in\{0,1\},
\end{equation}
over $\bbC^n$, with $\boldsymbol{\Sigma}_i\succ{\bf0}$. Then
\begin{equation}
p_i(\sfy)
=
\frac{1}{\pi^n\det(\boldsymbol{\Sigma}_i)}
\exp(-\sfy^\dagger\boldsymbol{\Sigma}_i^{-1}\sfy).
\end{equation}
Therefore,
\begin{align}
Z(p_0,p_1)
=
\frac{1}{
\sqrt{\det(\boldsymbol{\Sigma}_0)\det(\boldsymbol{\Sigma}_1)}
\det\left[(\boldsymbol{\Sigma}_0^{-1}+\boldsymbol{\Sigma}_1^{-1})/2\right]
}.
\end{align}
Using
\begin{equation}
\boldsymbol{\Sigma}_0^{-1}+\boldsymbol{\Sigma}_1^{-1}
=
\boldsymbol{\Sigma}_0^{-1}
(\boldsymbol{\Sigma}_0+\boldsymbol{\Sigma}_1)
\boldsymbol{\Sigma}_1^{-1},
\end{equation}
we recover~\eqref{eq:bhatt_complex_general_rewrite}.

For the occupancy problem,
\begin{equation}
\boldsymbol{\Sigma}_0=\sigma_{\rms}^{2}\bfI,
\qquad
\boldsymbol{\Sigma}_{1,\theta}
=
\sigma_{\rms}^{2}\bfI+\sigma_{\alpha_{\rms}}^2\bfg_\theta\bfg_\theta^\dagger.
\end{equation}
The matrix determinant lemma gives
\begin{equation}
\det(\boldsymbol{\Sigma}_{1,\theta})
=(\sigma_{\rms}^{2})^n(1+t_\theta)
\end{equation}
and
\begin{equation}
\det\left(\frac{\boldsymbol{\Sigma}_0+\boldsymbol{\Sigma}_{1,\theta}}{2}\right)
=(\sigma_{\rms}^{2})^n\left(1+\frac{t_\theta}{2}\right),
\end{equation}
which yields $z(t)=\sqrt{1+t}/(1+t/2)$.

\subsection{Convex Minorant and Covariance Relaxation}
\label{app:occupancy_covariance}
For
\begin{equation}
z(t)=\frac{2\sqrt{1+t}}{t+2},
\end{equation}
direct differentiation gives
\begin{equation}
z'(t)
=
-\frac{t}{\sqrt{1+t}(t+2)^2}
\label{eq:zprime_occ_appendix}
\end{equation}
and
\begin{equation}
z''(t)
=
\frac{3t^2-4}{2(1+t)^{3/2}(t+2)^3}.
\label{eq:zsecond_occ_appendix}
\end{equation}
Thus $z'$ decreases on $(0,2/\sqrt3)$ and increases thereafter. Let $t_0$ solve
\begin{equation}
z'(t_0)=\frac{z(t_0)-1}{t_0}.
\end{equation}
The unique positive solution is $t_0\simeq2.3830$, with $a_0\simeq-0.0674$. Define $g(t)=z(t)-(1+a_0t)$. Then $g(0)=g(t_0)=0$, while the shape of $z'$ implies that $g'$ changes sign once from positive to negative on $(0,t_0)$. Hence $g(t)\ge0$ on $[0,t_0]$. Since $t_0>2/\sqrt3$, $z$ is convex for $t\ge t_0$, and the line and $z$ have matching derivatives at $t_0$. Therefore the piecewise function in~\eqref{eq:def_und_z} is globally convex and lies below $z$.

Define
\begin{equation}
f_{\rm RDB}(u)
\triangleq
H_2^{-1}(1-\sqrt{1-u^2}),
\qquad0\le u\le1.
\end{equation}
Let $p(u)=f_{\rm RDB}(u)$ and $r(u)=1-\sqrt{1-u^2}$. Since $H_2(p(u))=r(u)$, while $H_2'(p)>0$ and $H_2''(p)<0$ for $0<p<1/2$, and $r'(u)\ge0$, $r''(u)>0$, implicit differentiation yields
\begin{equation}
p'(u)=\frac{r'(u)}{H_2'(p(u))}\ge0
\end{equation}
and
\begin{equation}
p''(u)
=
\frac{r''(u)-H_2''(p(u))[p'(u)]^2}{H_2'(p(u))}>0.
\end{equation}
Thus $f_{\rm RDB}$ is nondecreasing and convex. The TV outer map
\begin{equation}
f_{\rm TV}(u)=\frac12(1-\sqrt{1-u^2})
\end{equation}
is likewise nondecreasing and convex.

Starting from Proposition~\ref{prop:RDBoccupancy}, monotonicity gives
\begin{align}
P_e
&\ge
\bbE_{\bfX}
\left[
f_{\rm RDB}\left(
\bbE_\Theta[\underline z(t_\Theta(\bfX))]
\right)
\right].
\end{align}
Jensen's inequality for $f_{\rm RDB}$ and then for $\underline z$ yields
\begin{align}
P_e
&\ge
f_{\rm RDB}
\left(
\bbE_\Theta
\left[
\underline z\left(
\frac{\sigma_{\alpha_{\rms}}^2}{\sigma_{\rms}^{2}}
\bfv^\dagger(\Theta)\bfQ\bfv(\Theta)
\right)
\right]
\right)
\nonumber\\
&=
\Phi_{\rm RDB}(\bfQ),
\end{align}
where $\bfQ=\bbE[\bfX\bfX^\dagger]$. Replacing $f_{\rm RDB}$ by $f_{\rm TV}$ gives the TV covariance relaxation.

\section{Scalar MMSE Estimation for Circularly Symmetric Fading}
\label{app:nakagami_scalar_mmse}
Consider
\begin{equation}
\sfY=x\sfH+\sfZ,
\label{eq:scalar_fading_AWGN_appendix}
\end{equation}
where $x\in\bbC$ is known, $\sfZ\sim\cC\cN(0,\sigma_{\rms}^{2})$, and $\sfH=\sfR\rme^{j\Theta}$ with $\sfR$ and $\Theta$ independent and $\Theta\sim\operatorname{Unif}[0,2\pi)$. For $\ell\ge0$, define
\begin{equation}
g_\ell(a,b)
\triangleq
\bbE_{\sfR}\left[
\rme^{-a\sfR^2}\sfR^\ell I_\ell(2b\sfR)
\right],
\qquad a,b\ge0,
\label{eq:gell_appendix}
\end{equation}
where $I_\ell(\cdot)$ is the modified Bessel function of the first kind.

\begin{theorem}[Posterior mean for circularly symmetric fading]
\label{thm:scalar_MMSE_appendix}
For~\eqref{eq:scalar_fading_AWGN_appendix},
\begin{align}
\bbE[\sfH\mid\sfY=y,\sfX=x]
=
\frac{
g_1\left(\frac{|x|^2}{\sigma_{\rms}^{2}},\frac{|yx^*|}{\sigma_{\rms}^{2}}\right)
}{
g_0\left(\frac{|x|^2}{\sigma_{\rms}^{2}},\frac{|yx^*|}{\sigma_{\rms}^{2}}\right)
}
\rme^{j\angle(yx^*)}.
\label{eq:scalar_MMSE_estimator_appendix}
\end{align}
\end{theorem}

\begin{proof}
The likelihood is
\begin{equation}
f_{\sfY\mid\sfH,\sfX}(y\mid h,x)
=
\frac{1}{\pi\sigma_{\rms}^{2}}
\exp\left(-\frac{|y-xh|^2}{\sigma_{\rms}^{2}}\right).
\end{equation}
Writing $h=r\rme^{jt}$ gives
\begin{equation}
|y-xh|^2
=
|y|^2+|x|^2r^2-2|yx^*|r\cos(t-\angle(yx^*)).
\end{equation}
After marginalizing the phase,
\begin{align}
f_{\sfR\mid\sfY,\sfX}(r\mid y,x)
=
\frac{
f_{\sfR}(r)
\rme^{-\frac{|x|^2}{\sigma_{\rms}^{2}}r^2}
I_0\left(2\frac{|yx^*|}{\sigma_{\rms}^{2}}r\right)
}{
g_0\left(\frac{|x|^2}{\sigma_{\rms}^{2}},\frac{|yx^*|}{\sigma_{\rms}^{2}}\right)
}.
\label{eq:scalar_posterior_magnitude}
\end{align}
Conditioned additionally on $\sfR=r$, the phase is von Mises with location $\angle(yx^*)$, and its first circular moment is
\begin{equation}
\bbE[\rme^{j\Theta}\mid\sfY=y,\sfX=x,\sfR=r]
=
\frac{I_1\left(2\frac{|yx^*|}{\sigma_{\rms}^{2}}r\right)}
{I_0\left(2\frac{|yx^*|}{\sigma_{\rms}^{2}}r\right)}
\rme^{j\angle(yx^*)}.
\end{equation}
Averaging over~\eqref{eq:scalar_posterior_magnitude} proves the result.
\end{proof}

The resulting scalar MMSE is
{\small
\begin{align}
\operatorname{MMSE}(\sfH\mid\sfY,\sfX=x)
=
\bbE[|\sfH|^2]
-
\bbE\left[
\left(
\frac{
g_1\left(\frac{|x|^2}{\sigma_{\rms}^{2}},\frac{|\sfY x^*|}{\sigma_{\rms}^{2}}\right)
}{
g_0\left(\frac{|x|^2}{\sigma_{\rms}^{2}},\frac{|\sfY x^*|}{\sigma_{\rms}^{2}}\right)
}
\right)^2
\right].
\label{eq:scalar_MMSE_appendix}
\end{align}
}
For $\sfH\sim{\cC\text{Nakagami}}(m,\omega)$, $\bbE[|\sfH|^2]=\omega$. Related Bayesian estimators for circularly symmetric complex coefficients also lead naturally to Bessel-function representations~\cite{gerkmann2014bayesian}. The scalar expression illustrates why the exact nonlinear MIMO MMSE becomes analytically cumbersome; the required Nakagami radial integrals can be represented through confluent hypergeometric functions, but they do not yield a comparably simple full-MIMO MMSE formula. A complete exact-MMSE analysis of the MIMO Nakagami model is therefore left for future work.

\bibliographystyle{IEEEtran}
\bibliography{bibliography.bib}

@ARTICLE{7556344-Koch,
  author={Koch, Tobias},
  journal={IEEE Trans. Inf. Theory}, 
  title={The {S}hannon Lower Bound Is Asymptotically Tight}, 
  year={2016},
  volume={62},
  number={11},
  pages={6155-6161},
  doi={10.1109/TIT.2016.2604254}}

@ARTICLE{ISAC_Caire,
  author={Xiong, Yifeng and Liu, Fan and Cui, Yuanhao and Yuan, Weijie and Han, Tony Xiao and Caire, Giuseppe},
  journal={IEEE Trans. Inf. Theory}, 
  title={On the Fundamental Tradeoff of Integrated Sensing and Communications Under {G}aussian Channels}, 
  year={2023},
  volume={69},
  number={9},
  pages={5723-5751},
  doi={10.1109/TIT.2023.3284449}
}

@article{telatar1999capacity,
  title={Capacity of multi-antenna {G}aussian channels},
  author={Telatar, Emre},
  journal={European Trans. Telecommun.},
  volume={10},
  number={6},
  pages={585--595},
  year={1999},
  publisher={Wiley Online Library}
}

@article{goblick1965theoretical,
  title={Theoretical limitations on the transmission of data from analog sources},
  author={Goblick, T},
  journal={IEEE Trans. Inf. Theory},
  volume={11},
  number={4},
  pages={558--567},
  year={1965},
  publisher={IEEE}
}

@Book{Cover:2006,
author = {T. Cover and J. Thomas},
ALTeditor = {},
title = {Elements of Information Theory},
publisher = {New York:Wiley},
year = {2006},
OPTkey = {},
OPTvolume = {},
OPTnumber = {},
OPTseries = {},
OPTaddress = {},
edition = {2},
OPTmonth = {},
OPTnote = {},
OPTannote = {}
}

@inproceedings{riegler2018rate,
  title={Rate-distortion theory for general sets and measures},
  author={Riegler, Erwin and B{\"o}lcskei, Helmut and Koliander, G{\"u}nther},
  booktitle={IEEE Intern. Symp.  Inf. Theory (ISIT)},
  pages={101--105},
  year={2018},
  organization={IEEE}
}

@article{stam1959some,
  title={Some inequalities satisfied by the quantities of information of {F}isher and {S}hannon},
  author={Stam, Aart J},
  journal={Information and Control},
  volume={2},
  number={2},
  pages={101--112},
  year={1959},
  publisher={Elsevier}
}

@article{6g_isac,
  title={Integrated sensing and communications: Recent advances and ten open challenges},
  author={Lu, Shihang and Liu, Fan and Li, Yunxin and Zhang, Kecheng and Huang, Hongjia and Zou, Jiaqi and Li, Xinyu and Dong, Yuxiang and Dong, Fuwang and Zhu, Jia and others},
  journal={IEEE Internet Things J.},
  volume={11},
  number={11},
  pages={19094--19120},
  year={2024},
  publisher={IEEE}
}

@ARTICLE{WWbound,
  author={Weinstein, E. and Weiss, A.J.},
  journal={IEEE Trans. Inf. Theory}, 
  title={A general class of lower bounds in parameter estimation}, 
  year={1988},
  volume={34},
  number={2},
  pages={338-342},
  doi={10.1109/18.2647}}

@article{ziv1969some,
  title={Some lower bounds on signal parameter estimation},
  author={Ziv, Jacob and Zakai, Moshe},
  journal={IEEE Trans. Inf. Theory},
  volume={15},
  number={3},
  pages={386--391},
  year={1969},
  publisher={IEEE}
}

@ARTICLE{Ziv-ZakBOund,
  author={Jeong, Minoh and Dytso, Alex and Cardone, Martina},
  journal={IEEE Trans. Inf. Theory}, 
  title={A Comprehensive Study on {Z}iv-{Z}akai Lower Bounds on the {MMSE}}, 
  year={2025},
  volume={71},
  number={4},
  pages={3214-3236},
  doi={10.1109/TIT.2025.3541987}}

@article{dytso2023meta,
  title={Meta derivative identity for the conditional expectation},
  author={Dytso, Alex and Cardone, Martina and Zieder, Ian},
  journal={IEEE Trans. Inf. Theory},
  volume={69},
  number={7},
  pages={4284--4302},
  year={2023},
  publisher={IEEE}
}

@inproceedings{aras2019family,
  title={A family of {B}ayesian {C}ram{\'e}r-{R}ao bounds, and consequences for log-concave priors},
  author={Aras, Efe and Lee, Kuan-Yun and Pananjady, Ashwin and Courtade, Thomas A},
  booktitle={IEEE Intern. Symp.  Inf. Theory (ISIT)},
  pages={2699--2703},
  year={2019},
  organization={IEEE}
}

@article{dytso2019mmse,
  title={ {MMSE} bounds for additive noise channels under {K}ullback--{L}eibler divergence constraints on the input distribution},
  author={Dytso, Alex and Fau{\ss}, Michael and Zoubir, Abdelhak M and Poor, H Vincent},
  journal={IEEE Trans. Sig. Proc.},
  volume={67},
  number={24},
  pages={6352--6367},
  year={2019},
  publisher={IEEE}
}

@article{miller1978modified,
  title={A modified {C}ram{\'e}r-{R}ao bound and its applications (Corresp.)},
  author={Miller, R and Chang, Chow},
  journal={IEEE Trans. Inf. Theory},
  volume={24},
  number={3},
  pages={398--400},
  year={1978},
  publisher={IEEE}
}

@ARTICLE{Zamir-Tightness,
  author={Linder, T. and Zamir, R.},
  journal={IEEE Trans. Inf. Theory}, 
  title={On the asymptotic tightness of the {S}hannon lower bound}, 
  year={1994},
  volume={40},
  number={6},
  pages={2026-2031},
  doi={10.1109/18.340474}}

@ARTICLE{smartcity,
  author={Zhu, Xiaoqiang and Liu, Jiqiang and Lu, Lingyun and Zhang, Tao and Qiu, Tie and Wang, Chunpeng and Liu, Yuan},
  journal={IEEE Commun. Surveys  Tuts.}, 
  title={Enabling Intelligent Connectivity: A Survey of Secure {ISAC} in {6G} Networks}, 
  year={2025},
  volume={27},
  number={2},
  pages={748-781},
  doi={10.1109/COMST.2024.3432871}}

@ARTICLE{isac_6g_multifunc,
  author={Wei, Zhongxiang and Liu, Fan and Masouros, Christos and Su, Nanchi and Petropulu, Athina P.},
  journal={IEEE Commun. Mag.}, 
  title={Toward Multi-Functional {6G} Wireless Networks: Integrating Sensing, Communication, and Security}, 
  year={2022},
  volume={60},
  number={4},
  pages={65-71},
  doi={10.1109/MCOM.002.2100972}}

@INPROCEEDINGS{V2X_mizmizi,
  author={Dong, Kai and Mizmizi, Marouan and Tagliaferri, Dario and Spagnolini, Umberto},
  booktitle={IEEE Int. Conf.  Commun.}, 
  title={Vehicular Blockage Modelling and Performance Analysis for mm{W}ave {V2V} Communications}, 
  year={2022},
  volume={},
  number={},
  pages={3604-3609},
  doi={10.1109/ICC45855.2022.9838711}}

@book{Tsybakov2009Nonparametric,
  author    = {Tsybakov, Alexandre B.},
  title     = {Introduction to Nonparametric Estimation},
  publisher = {Springer},
  address   = {New York, NY},
  year      = {2009},
  series    = {Springer Series in Statistics},
  isbn      = {978-0-387-79051-0},
  doi       = {10.1007/b13794}
}

@ARTICLE{alouini1999capacity,
  author={Alouini, M.-S. and Goldsmith, A.J.},
  journal={IEEE Transactions on Vehicular Technology}, 
  title={Capacity of {R}ayleigh fading channels under different adaptive transmission and diversity-combining techniques}, 
  year={1999},
  volume={48},
  number={4},
  pages={1165-1181},
  doi={10.1109/25.775366}
  }

@article{Liu2022JSAC,
  author  = {F. Liu and Y. Cui and C. Masouros and J. Xu and T. X. Han and Y. C. Eldar and S. Buzzi},
  title   = {Integrated Sensing and Communications: Toward Dual-Functional Wireless Networks for {6G} and Beyond},
  journal = {IEEE J. Sel. Areas Commun.},
  volume  = {40},
  number  = {6},
  pages   = {1728--1767},
  year    = {2022}
}

@ARTICLE{HuaHanXu2024,
  author={Hua, Haocheng and Han, Tony Xiao and Xu, Jie},
  journal={IEEE Transactions on Wireless Communications}, 
  title={{MIMO} Integrated Sensing and Communication: {CRB}-Rate Tradeoff}, 
  year={2024},
  volume={23},
  number={4},
  pages={2839-2854},
  doi={10.1109/TWC.2023.3303326}
  }

@inproceedings{alouini1997capacity,
  title={Capacity of {N}akagami multipath fading channels},
  author={Alouini, M-S and Goldsmith, Andrea},
  booktitle={IEEE 47th Veh. Technol. Conf.},
  volume={1},
  pages={358--362},
  year={1997},
  organization={IEEE}
}

@article{Blahut1972IT,
  author  = {Blahut, Richard E.},
  title   = {Computation of Channel Capacity and Rate-Distortion Functions},
  journal = {IEEE Trans. Inf. Theory},
  volume  = {18},
  number  = {4},
  pages   = {460--473},
  year    = {1972}
}

@article{Arimoto1972IT,
  author  = {Arimoto, Shizuo},
  title   = {An Algorithm for Computing the Capacity of Arbitrary Discrete Memoryless Channels},
  journal = {IEEE Trans. Inf. Theory},
  volume  = {18},
  number  = {1},
  pages   = {14--20},
  year    = {1972}
}

@ARTICLE{ahmadipour2024information,
  author={Ahmadipour, Mehrasa and Kobayashi, Mari and Wigger, Michele and Caire, Giuseppe},
  journal={IEEE Transactions on Information Theory}, 
  title={An Information-Theoretic Approach to Joint Sensing and Communication}, 
  year={2024},
  volume={70},
  number={2},
  pages={1124-1146},
  doi={10.1109/TIT.2022.3176139}
  }

@article{gerkmann2014bayesian,
  title={Bayesian estimation of clean speech spectral coefficients given a priori knowledge of the phase},
  author={Gerkmann, Timo},
  journal={IEEE Trans. Signal Process.},
  volume={62},
  number={16},
  pages={4199--4208},
  year={2014},
  publisher={IEEE}
}

@inproceedings{liu2023deterministic,
  title={Deterministic-random tradeoff of integrated sensing and communications in {G}aussian channels: A rate-distortion perspective},
  author={Liu, Fan and Xiong, Yifeng and Wan, Kai and Han, Tony Xiao and Caire, Giuseppe},
  booktitle={IEEE Intern. Symp.  Inf. Theory (ISIT)},
  pages={2326--2331},
  year={2023}
}

@ARTICLE{Fei24KL,
  author={Fei, Zesong and Tang, Shuntian and Wang, Xinyi and Xia, Fanghao and Liu, Fan and Andrew Zhang, J.},
  journal={IEEE Wireless Communications Letters}, 
  title={Revealing the Trade-Off in {ISAC} Systems: The {KL} Divergence Perspective}, 
  year={2024},
  volume={13},
  number={10},
  pages={2747-2751},
  doi={10.1109/LWC.2024.3443409}
  }

\end{document}